\documentclass[]{revtex4}
\usepackage{graphicx}
\usepackage{psfrag}

\newcommand{\be}{\begin{equation}}
\newcommand{\ee}{\end{equation}}
\newcommand{\beq}{\begin{eqnarray}}
\newcommand{\eeq}{\end{eqnarray}}
\begin{document}

\title{A Gaussian-optical Approach to Stable Periodic Orbit
Resonances of Partially Chaotic Dielectric Micro-cavities}

\author{H.~E.~Tureci, H.~G.~L. Schwefel and A.~Douglas~Stone}
\affiliation{Department of Applied Physics, P. O. Box 208284, Yale University,
New Haven, CT 06520-8284}
\email{hakan.tureci@yale.edu}
\date{June 26, 2002}
\author{E.~E.~Narimanov}
\affiliation{Department of Electrical Engineering, Princeton University, Princeton, NJ 08544, USA}
\begin{abstract}
The quasi-bound modes localized on stable periodic ray orbits of
dielectric micro-cavities are constructed in the short-wavelength
limit using the
parabolic equation method. These modes are shown to coexist with irregularly
spaced ``chaotic" modes for the generic case. The
wavevector quantization rule for the quasi-bound modes is derived and
given a simple
physical interpretation in terms of Fresnel reflection; quasi-bound modes are
explictly constructed and compared to numerical results.  The effect of
discrete symmetries of the resonator is analyzed and shown to give rise
to quasi-degenerate multiplets; the average splitting of these multiplets
is calculated by methods from quantum chaos theory.
\end{abstract}

\maketitle

\section{Introduction}

There is currently a great deal of interest in dielectric
micro-cavities which can serve as high-Q
resonators by confining light on the basis of multiple reflections
from a boundary of dielectric
mismatch \cite{campillo_book,yokoyama1,nobel,science,nature}. Such 
resonators have been
used to study fundamental optical physics such as cavity
quantum electrodynamics \cite{campillo1,lin1} and have been proposed 
and demonstrated as
the basis of both active and
passive optical components \cite{vahala1,little1}.  Of particular 
interest in this work are
dielectric micro-cavity
lasers which have already been demonstrated for a wide variety
of shapes: spheres\cite{chang3}, cylinders \cite{campillo_book},
squares \cite{poon1}, hexagons \cite{braun1}, and deformed cylinders 
and spheres (asymmetric
resonant cavities \cite{nobel} - ARCs)\cite{science,mekis,nockel1,gornik1,chang1,rex1}.
The work on ARC micro-cavity lasers has shown the possibility of producing high
power directional emission from
such lasers, which lase in different spatial mode patterns depending
on the index of refraction
and precise shape of the boundary. For example modes based on stable
periodic ray orbits with bow-tie \cite{science} and
triangular  geometry \cite{rex_thesis} have been observed, as well as 
whispering
gallery-like modes \cite{chang1,rex_thesis} which are not obviously
related  to any periodic ray orbit.  In addition a periodic orbit
mode can be selected for lasing even
if it is unstable \cite{rex1,gmachl1,sblee1}; due to the analogy to quantum 
wavefunctions based
on unstable classical periodic
orbits, such modes have been termed ``scarred'' \cite{heller1}. 
There is at present
no quantitative understanding
of the mode selection mechanism in ARCs and theory has tended to work
backwards from experimental
observations; however the passive cavity solutions have been found
to explain quite well the observed lasing emission patterns.
The formal analogy between ARCs and the problem of classical and
quantum billiards has
given much insight into their emission properties 
\cite{campillo_book,nobel}.  For example it
was predicted and recently
observed that polymer ARCs of elliptical shape (index $n=1.49$) have
dramatically different emission
patterns from quadrupolar shaped ARCs with the same major to minor 
axis ratio \cite{harald1}.
The difference can be fully understood by the different structure of
the phase space for ray motion in
the two cases, the ellipse giving rise to integrable ray motion and
the quadrupole to partially chaotic
(mixed) dynamics.  Resonators are of course open systems
for which radiation can leak out to
infinity.  In the discussion of the ray-wave correspondence
immediately following we neglect this leakage assuming
only perfect specular reflection of light rays at the dielectric
boundary.  However the theory of ARCs \cite{nobel,nature} includes
these effects and they will be treated when relevant below.

\begin{figure}[h]
\centering
\includegraphics[width=\linewidth]{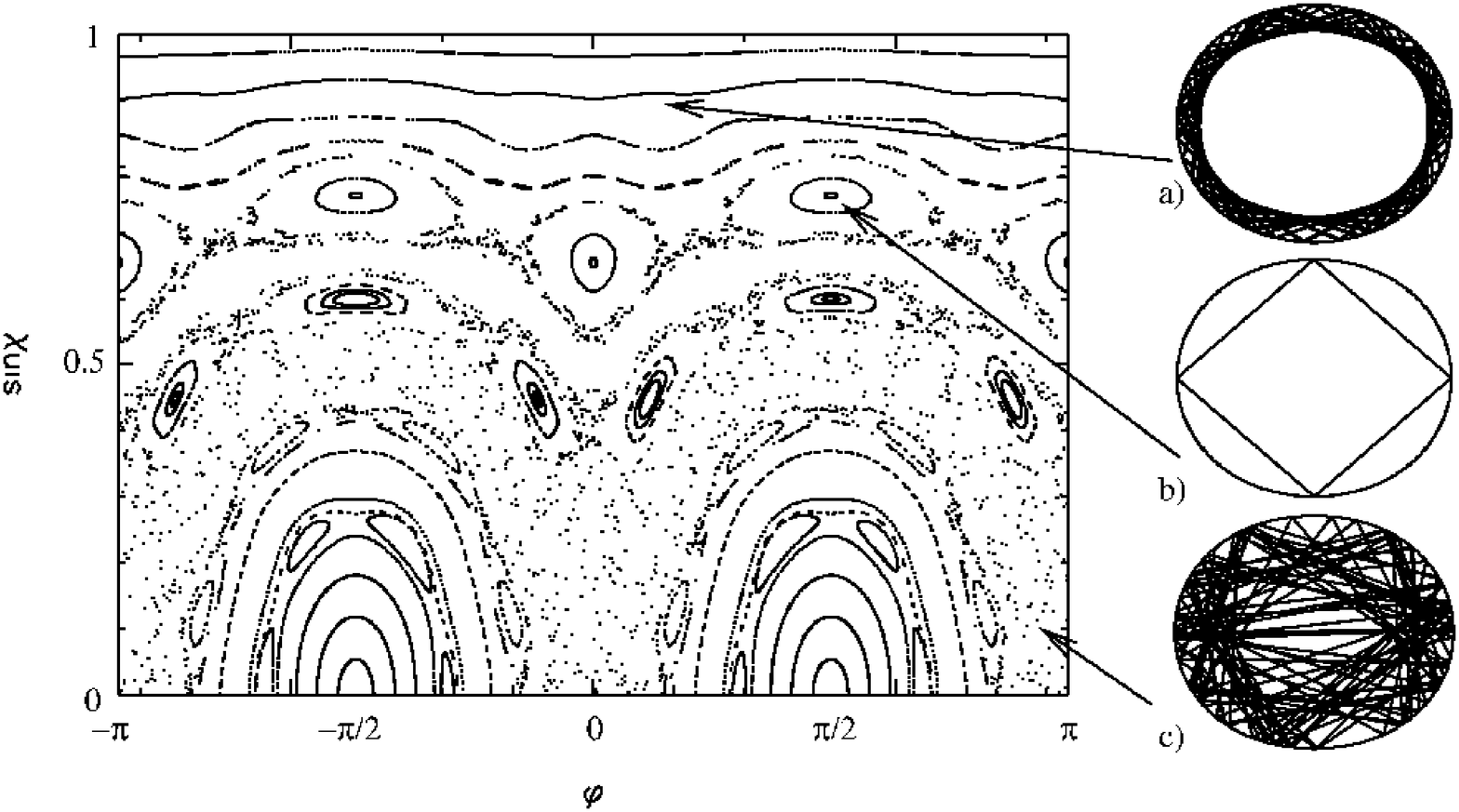}
\caption{Surface of section illustrating the different regions of 
phase space for a
closed quadrupole billiard with boundary given by $r(\phi) = R(1 + 
\epsilon \cos
2 \phi)$ for $\epsilon = 0.072$.  Real-space ray trajectories 
corresponding to each
region are indicated at left: a) A quasi-periodic, marginally stable orbit.
  b) A  stable four-bounce ``diamond'' periodic orbit (surrounded by stability
``islands'' in the SOS) c) A chaotic ray trajectory.  Orbits of type (b)
have associated with them regular gaussian solutions as we will show below.}
\label{fig:billiard}
\end{figure}

As is well-known, the geometric optics of a uniform dielectric region
with perfectly reflecting
boundaries is formally analogous to the problem of a point mass
moving in a billiard and hence
we may use the terminology of Hamiltonian dynamics to describe ray
motion in the resonator.  As discussed below, we shall specialize to
cylindrical geometries in which the relevant motion will be in the
the plane transverse to the axis, hence we focus on two-dimensional
billiards.  The quadrupole billiard is an example of a generic
deformation of the
circle or cylinder in that it is smooth and analytic and does not preserve any
constant of motion.  (Motion in
the circle of course conserves angular momentum and motion in a closed
elliptical cavity conserves a generalized angular
momentum, the product of the instantaneous angular momenta with
respect to each focus \cite{berry2}).  A generic
deformation such as the quadrupole will lead to a phase space for ray
motion which has three types
of possible motion (depending on the choice of initial conditions):
oscillatory motion in the vicinity of
a stable periodic ray orbit, chaotic motion in regions associated
with unstable periodic ray orbits,
and marginally stable motion associated with families of
quasi-periodic orbits (motion on a
so-called KAM torus). To elucidate the structure of phase space it is
conventional to plot a number of representative trajectories in a 
two-dimensional
cut through phase space, called the surface of section 
(Fig.~\ref{fig:billiard}).
For our system the surface of section corresponds to the boundary of 
the billiard
and the coordinates of the ray are the polar angle $\phi$ and the 
angle of incidence
$\sin \chi$ at each bounce.  The three types of motion described above
are illustrated for the quadrupole billiard in Fig.~\ref{fig:billiard} both in
phase space and in real space.

It has been shown \cite{robnik1} that the solutions of the wave 
equation for a generic shape
such as the quadrupole can be classified by their association
with these three different kinds of
motion. The ray-mode (or wave-particle) correspondence becomes stronger
as we approach the short-wavelength
(semi-classical) limit, which in this work is defined by $kl \gg 1$
where $k$ is the wavevector and $l$ is a
typical linear dimension of the resonator, e.g. the average radius. The
modes associated with quasi-periodic families can be treated
semiclassically by eikonal methods of the  type
introduced, e.g. by Keller
\cite{keller1}, and referred to in its most general form as EBK
(Einstein-Brillouin-Keller) quantization.  The
individual modes associated with unstable periodic orbits and chaotic
motion cannot be treated by any current
analytic methods (although the density of states for a chaotic system
can be found by a sophisticated analytic
method based on Gutzwiller's Trace Formula \cite{gutz_book}).
Finally, the modes associated with stable periodic orbits can be
treated by generalizations of gaussian optics and
will be the focus of the current work.

\begin{figure}[h]
\centering
\includegraphics[width=\linewidth]{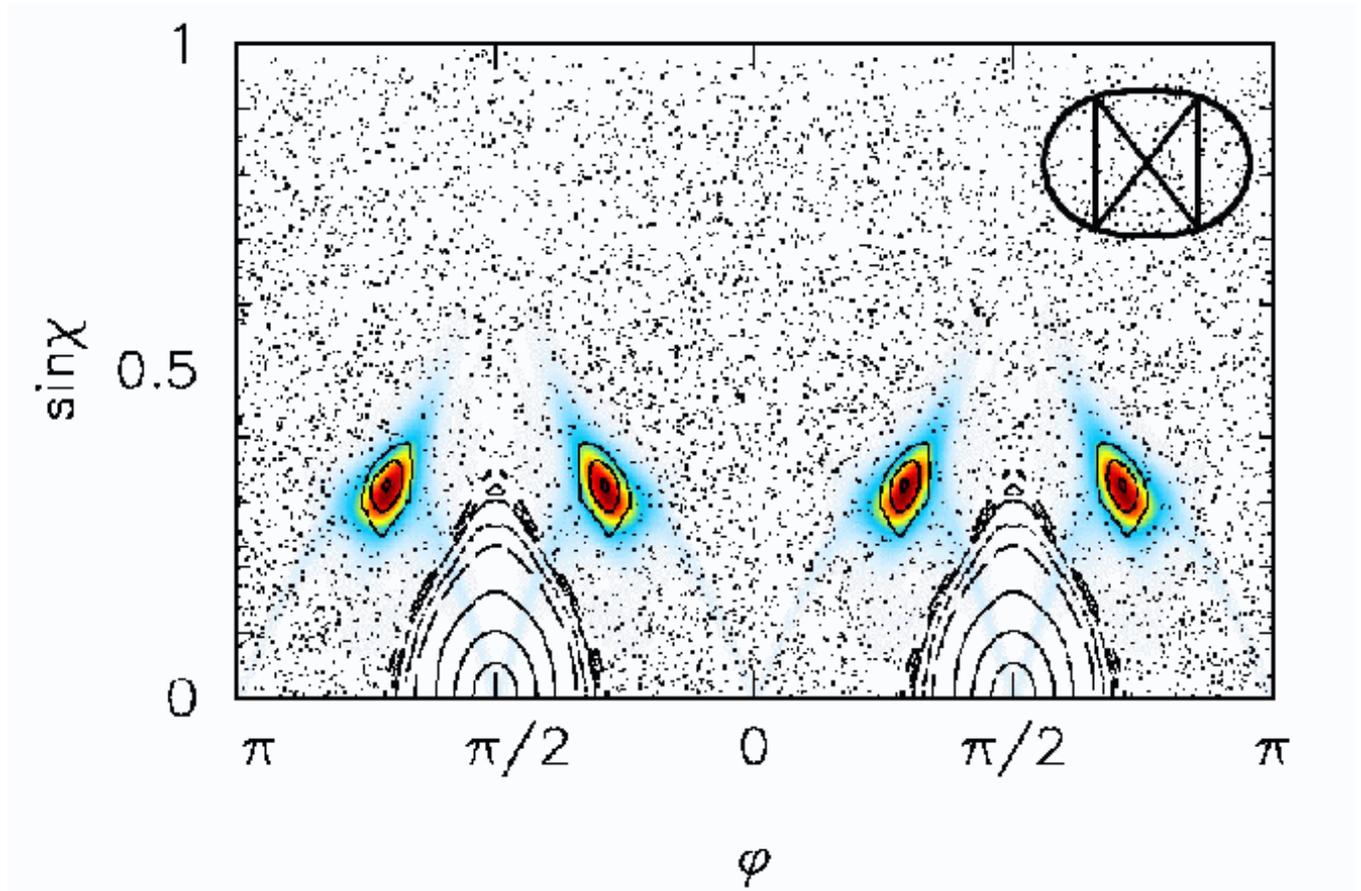}
\caption{Black background gives the surface of section for the quadrupole
at $\epsilon = 0.17$ for which the four small islands correspond to a
stable bow-tie shaped orbit (inset). A numerical solution of the Helmholtz
equation for this resonator can be projected onto this surface of 
section via the Husimi transform [27] and is found to have high intensity
(in false color scale) precisely on these islands, indicating that this is a mode 
associated with the bow-tie orbit.}
\label{fig:husimi}
\end{figure}

If a mode is found by numerical solution, its
interpretation in terms of the ray phase space can be determined with
reasonable accuracy by means of the
Husimi projection onto the phase space (see Fig.~\ref{fig:husimi}), 
although it is
well-known that for $kl$ not much greater
than unity the exact solutions tend to smear out in the phase
space over regions of order $1/kl$
and do not correspond very closely to specific classical structures.
For a closed generic ARC the full spectrum will look highly irregular
(see Fig.~\ref{fig:spectrum}(a)), but contained in the full spectrum 
will be regular
sequences associated with tori and stable periodic orbits 
(Fig.~\ref{fig:spectrum}(b)).
The stable periodic orbit modes will give
the simplest such sequences consisting of two different constant
spacings, one associated with the
longitudinal quantization of the orbit (free spectral range) and the
other associated with transverse
excitations. In the example of Fig.~\ref{fig:spectrum}
the imbedded regular spectrum is due to the stable ``bow-tie''
orbit. The regular portion of the spectrum is extracted by weighting each
level by the overlap of its Husimi function with the islands corresponding
to the stable periodic orbit in the surface of section.  Clearly, hidden
within this complex spectrum are simple regular mode sequences of the
type familiar from Gaussian optics.  In the current work we show how to
calculate the resonant energies and spatial
intensity patterns of such modes associated with arbitrary stable
periodic ray orbits
for both the ideal closed resonator
and a dielectric resonator of the same shape with arbitrary
dielectric mismatch $n$.  We shall
refer to these as periodic orbit modes or PO modes.

In the case of the open resonator, the modes have a
width which can be expressed
as a negative imaginary part of k, and some of the PO modes may be so
broad (short-lived)
that they would not appear as sharp spectral lines.
Within the  gaussian-optical
theory we present below that this width is entirely determined by Fresnel
reflection at the interface
and would be zero for a periodic orbit which has all bounces above
the total internal reflection
condition, but would be quite large for a periodic orbit, such as the
two-bounce Fabry-Perot orbit,
which has normal incidence on the boundary. An exact solution must find a
non-zero width for {\it all} PO modes, due to evanescent leakage across a
curved interface, even if all the bounces
satisfy the total internal reflection condition.

\begin{figure}[h]
\centering
\includegraphics[width=0.7\linewidth]{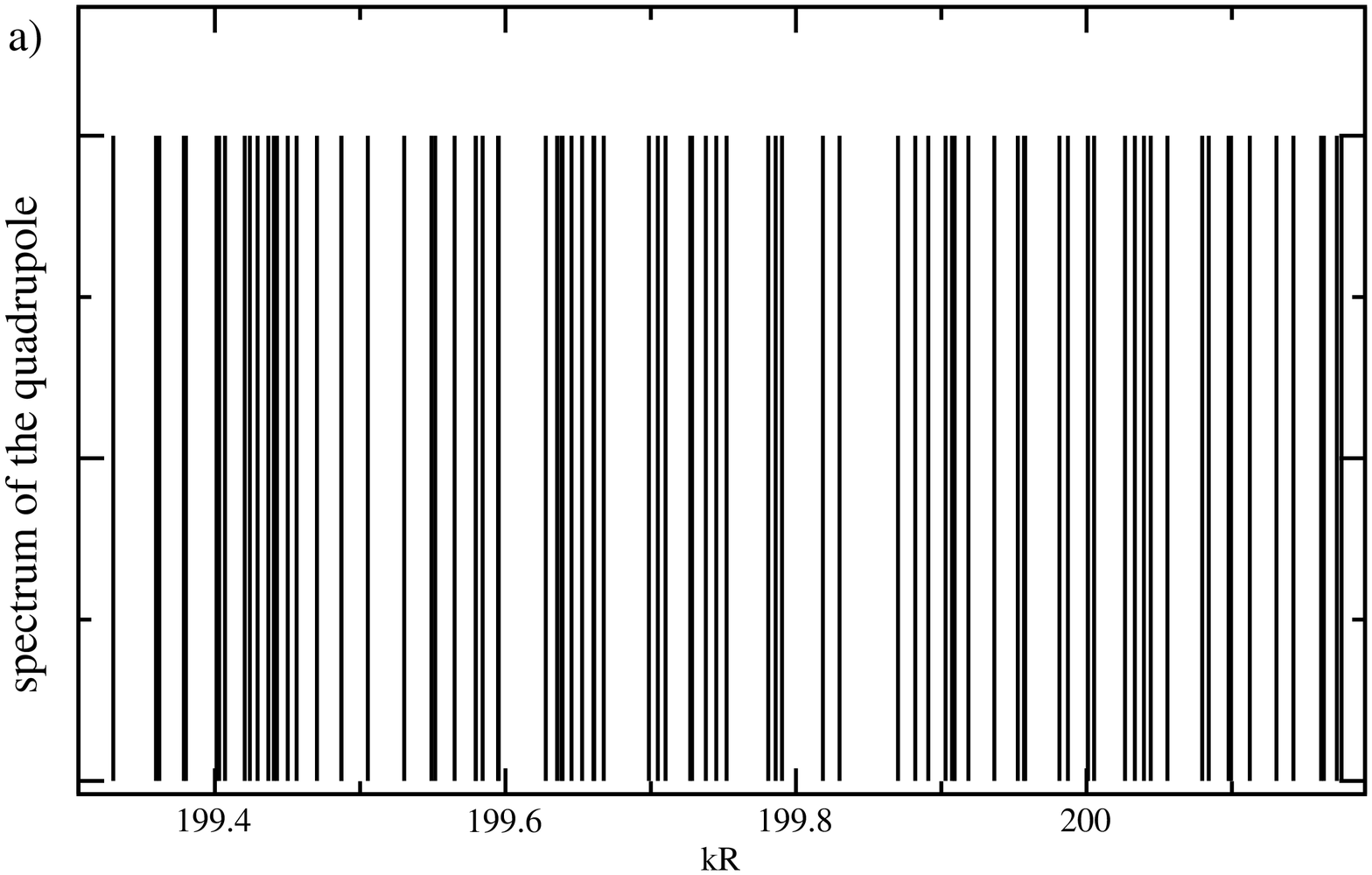}
\end{figure}
\begin{figure}[h]
\centering
\includegraphics[width=0.7\linewidth]{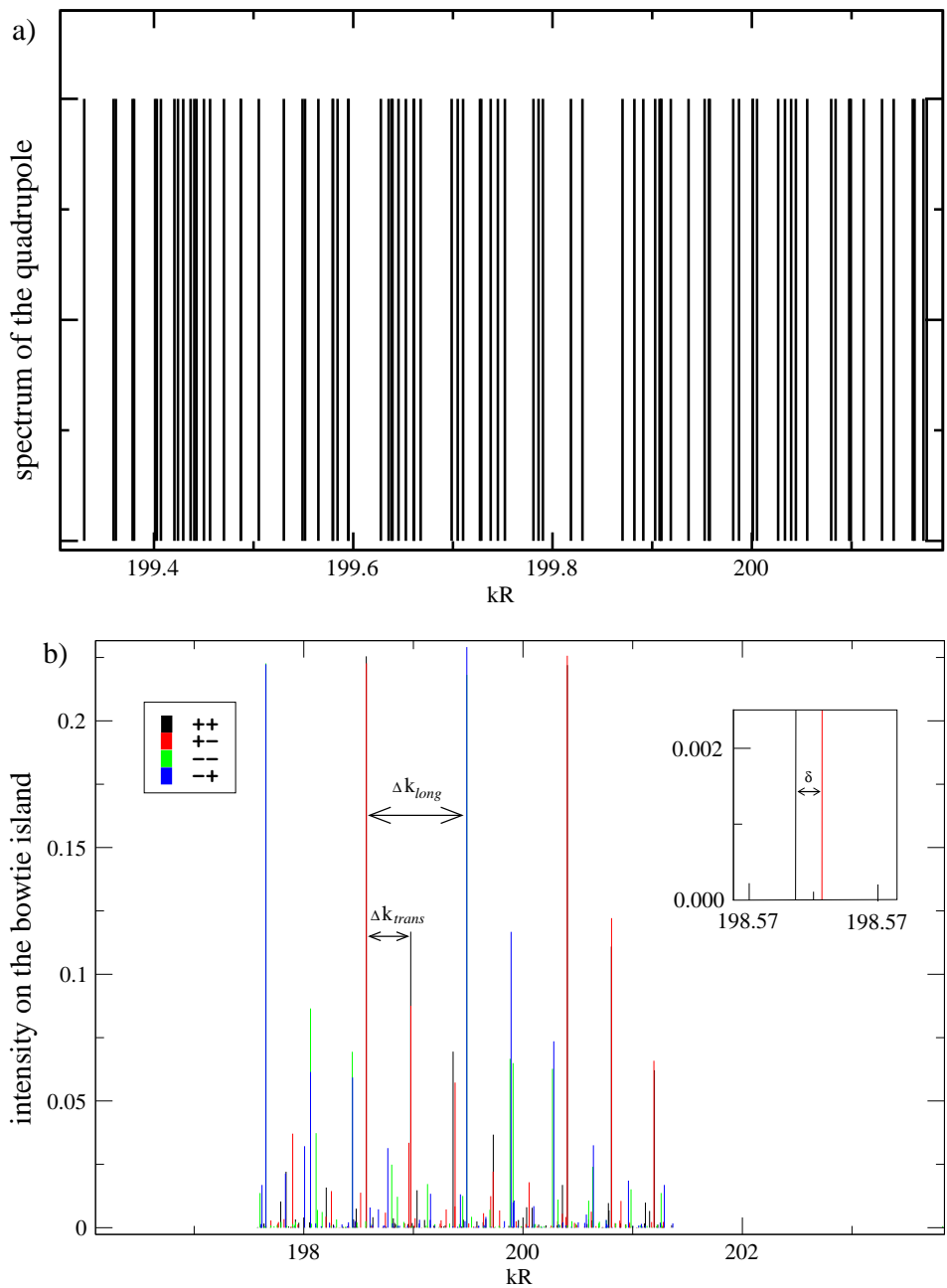}
\caption{(a) Vertical lines indicate wavevectors of bound states of the closed
quadrupole resonator for $\epsilon = 0.17$; no regular spacings are
visible.  (b) Spectrum weighted by overlap of the Husimi function of the
solution with the bow-tie island as in Fig.~\ref{fig:husimi}.  Note the emergence of
regularly spaced levels with two main spacings $\Delta k_{long}$ and
$\Delta k_{trans}$.  These spacings, indicated by the arrows, are
calculated from the length of the bow-tie orbit and the associated
Floquet phase (see section \ref{section:quant1} below).  The color coding corresponds to
the four possible symmetry types of the solutions (see section \ref{section:sym1} below).
In the inset is a magnified view showing the splitting of quasi-degenerate
doublets as discussed in section \ref{section:split}. Note the pairing of the $(+,+)$ 
and $(+,-)$ symmetry types as predicted in section \ref{section:sym2}. The different symmetry
pairs alternate every free spectral range ($\Delta k_{long}$).}
\label{fig:spectrum}
\end{figure}

Another limitation of the gaussian theory of stable PO modes is that
it predicts exactly degenerate modes
when the associated orbit has discrete symmetries, even in cases for
which a group-theoretic analysis
shows that there can be no exact symmetries (this is
the case, for example in the quadrupole).
Instead the exact solutions will have some integer {\it
quasi-degeneracy} in which the spectrum consists of nearly
degenerate multiplets, whose multiplicity depends in detail on the
particular PO mode. This point is illustrated by the inset to 
Fig.~\ref{fig:spectrum}(b).
We will show
below how to calculate the multiplicity of these quasi-degeneracies
for a given PO
and introduce a theoretical approach to
estimate the size of the associated splittings.

\section{Gaussian optical approach to the closed cavity}

\begin{figure}[h]
\psfrag{bb}{$\partial D$}
\centering
\includegraphics[width=0.6\linewidth]{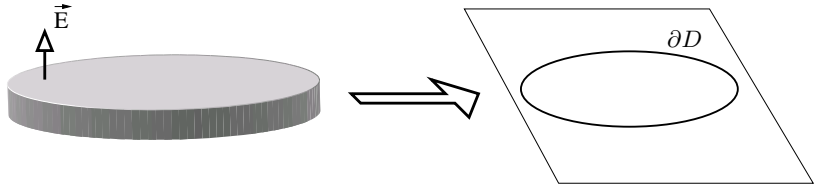}
\caption{Illustration of the reduction of the Maxwell equation for an
infinite dielectric cylinder to the 2D Helmholtz equation for the
TM case (E field parallel to axis) and $k_z=0$.}
\label{figcylinder}
\end{figure}

The quantization of electromagnetic modes within dielectric bodies
enclosed by metallic boundaries,
including the case of arbitrary
index variation inside the resonator and three-dimensions has been
treated before in the literature
\cite{babic_book}.  However the generality of the treatment makes it
difficult to extract simple results
of use to researchers working with uniform dielectric optical
micro-cavities; more importantly all the
work of which we are aware focuses on the case of Dirichlet boundary
conditions corresponding to
perfect reflection at the boundary. Perfect reflection of course
leads to true bound states. In the next
section we show how to generalize these results to the correct
boundary conditions at a dielectric interface
and hence for the case of quasi-bound as opposed to bound states.
However, first, in this section we
develop the formalism for the closed case which we will generalize to
the case of interest.  In all of this work we will specialize to the
case of two
dimensions, corresponding to
an infinite dielectric cylinder with an
arbitrary cross-section in the transverse
plane (see Fig.~\ref{figcylinder}) with the condition $k_z=0$.  To conform with
conventions introduced below we will refer to the two-dimensional
coordinate system as $(X,Z)$.
In this case the TM and TE polarizations
separate and we have a scalar wave equation with simple continuity
conditions at the boundary for
the electric field (for TM) or magnetic field (for TE).  For the
closed case (for which the field is zero on the boundary) we can set
the index $n=1$
and work simply with the Helmholtz equation.

Consider the solutions $E(X,Z)$ of the Helmholtz equation in two dimensions:
\be
\left( \nabla^2 + k^2 \right) E = 0
\label{eqhh}
\ee
This solution is assumed to be defined in a bounded two dimensional domain $D$,
with a boundary $\partial D$ on which $E=0$, leading to discrete
real eigenvalues $k$.  We are interested in
a subset of these solutions \emph{for asymptotically large values of}
$k$ for which the  eigenfunctions are localized around stable
periodic orbits(POs) of the specularly reflecting boundary $\partial D$.

\subsection{The Parabolic equation approximation}
\begin{figure}[h]
\psfrag{bb}{$\partial D$}
\centering
\includegraphics[width=0.5\linewidth]{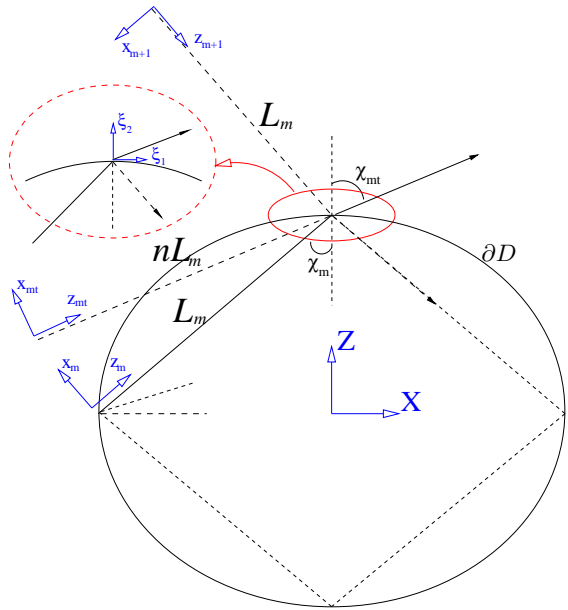}
\caption{Coordinate system and variables used in the text displayed
for the case of a quadrupolar boundary $\partial D$ and the diamond
four-bounce PO. A fixed coordinate system $(X,Z)$
is attached to the origin. The ``mobile'' coordinate systems 
$(x_m,z_m)$ are fixed
on segments of the periodic orbit so that their respective $z$-axes 
are parallel to
the segment, while their origins are set back a distance $L_m$ (or $nL_m$ for
transmitted beam axes), so as to account for zeroth order phase accumulation
between successive bounce-points. $\xi_1,\xi_2$ are the common local 
coordinates
at each bounce (index $m$ suppressed). Scaled coordinates are
denoted by tildes, e.g. $\tilde{x}_m=\sqrt{k}x_m$. The coordinate 
transformations
at each bounce $m$ are given by $z_i = L_m + \xi_1 \sin\chi_i + \xi_2 
\cos\chi_i$,
$z_r = L_m + \xi_1 \sin\chi_i - \xi_2 \cos\chi_i$, $z_{t} = nL_m + 
\xi_1 \sin\chi_t
+ \xi_2 \cos\chi_t$ and $x_i = \xi_1 \cos\chi_i - \xi_2 \sin\chi_i$, 
$x_{r} = \xi_1
\cos\chi_i + \xi_2 \sin\chi_i$, $x_{t} = \xi_1 \cos\chi_{t} - \xi_2 
\sin\chi_{t}$,
where $i,r,t$ refer to the incident, transmitted and reflected solutions.}
\label{figcs}
\end{figure}

The ``N-bounce PO''s corresponding to a boundary $\partial D$ are the
set of ray orbits which close upon
themselves upon reflecting specularly N times.  The shape of the
boundary defines a non-linear map from the incident angle and polar
angle at the $m^{th}$ bounce $(\phi_m,\sin \chi_m)$ to that at the 
$m+1$ bounce.
Typical trajectories of this map are shown in the surface of section plot of
Fig. 1.  The period-N orbits are the fixed points of the $N^{th}$
iteration of this map.  For a given period-N orbit (such as the period
four ``diamond'' orbit shown in Fig.~\ref{figcs}),
let the length of $m$th segment (``arm") be $l_m$, the accumulated
distance from origin be $L_M=\sum_{m=1}^{M}~l_m$, and
$L=L_N$ be the length of the entire PO. We are looking for modal
solutions which
are localized around the PO and decay in the transverse direction, hence we
express Eq.~(\ref{eqhh}) in Cartesian coordinates $(x_m , z_m)$
attached to the PO, where $z_m$-axis is aligned
with $m$th arm and $x_m$ is the transverse coordinate.
We also use $z$ to denote the cumulative length along the PO, which varies
in the interval $( -\infty, +\infty )$.

We write the general solution as:
\be
E(X,Z) = \sum_{m=1}^{N} E_m (x_m(X,Z), z_m(X,Z))
\label{eqans1}
\ee
where the ``local set" $(x_m,z_m)$ and the fixed set $(X,Z)$ are
related by shifts and rotations
(see Fig.~\ref{figcs}). Next, in accordance with the
parabolic equation approximation \cite{babic_book},
we assume that the main variation of the phase in z-direction is
linear (``slowly
varying envelope approximation") and factor it out:
\be
E_m (x_m, z_m) = u_m(x_m,z_m) e^{ikz_m}
\label{eqans2}
\ee
This definition suggests defining the origins of the local
coordinates such that $z_m=z$ along the PO, and we do so (see
Fig.~\ref{figcs}).

Next, we insert the solution Eq.~(\ref{eqans1}) into Eq.~(\ref{eqhh}) and
using the invariance of the Laplacian, we obtain:
\be
\left( \nabla_m^2 + k^2 \right) E_m = 0
\label{eqlo}
\ee
and the boundary condition translates into
\be
E_m + E_{m+1} |_{\partial D} = 0
\label{eqbc}
\ee
Here $\nabla_m$ is the Laplacian expressed in local coordinate system.
This reduction is possible as long as the solutions are
well-localized, and the bounce points of the PO are well-separated
(with respect to $k^{-1}$), even if the PO were to self-intersect.
These assumptions will be justified by the ensuing
construction.

Dropping for the moment the arm index, we will focus on
Eq.~(\ref{eqlo}). Inserting Eq.~(\ref{eqans2}), we arrive at
\be
u_{xx} + u_{zz} + 2ik u_z = 0
\label{eqred1}
\ee
The basic assumption of the method is that after removing phase
factor $\exp [ikz]$ which varies on the
scale of the wavelength
$\lambda$, the $z$-dependence of $u$ is slow, i.e. $u_z \sim
u/l,u_{zz} \sim u/l^2$, where $l$ is a typical linear
dimension associated with the boundary, e.g. a chord length of the
orbit or the curvature at a bounce point. In the
semiclassical limit $ l \gg \lambda$.  The transverse ($x$) variation
of $u$ on the other hand is assumed to occur
on a scale $\sqrt{l\lambda}$, intermediate between the wavelength and
the cavity scale, $l$; hence the
transverse variation of $u(x,z)$ is much more rapid than its
longitudinal variation.  This motivates us to
introduce the scaling $\tilde{x}=\sqrt{k}x$ to treat this boundary
layer, which leads to
\be
u_{\tilde{x}\tilde{x}} + \frac{1}{k} u_{zz} + 2i u_z = 0.
\ee
We can now neglect the $u_{zz}$ term due to the condition $kl \gg 1$, and
obtain a partial differential equation of parabolic type:
\be
{\cal L}\; u(\tilde{x},z) = 0
\label{eqred2}
\ee
where ${\cal L}=\partial_{\tilde{x}}^2 + 2i\partial_z$. Next, we make
the ansatz
\be
u(x,z) = c \: A(z) \exp{\left[\frac{i}{2}\Omega(z)\tilde{x}^2\right]}
\label{eqans3}
\ee
Inserting Eq.~(\ref{eqans3}) into Eq.~(\ref{eqred2}) and requiring it
to be satisfied for all $x$, we obtain the relations
\beq
\Omega^2 + \Omega^{\prime} & = & 0 \\
A\Omega + 2 A^{\prime} & = & 0
\eeq
Here and in the rest of the text we will use primes as a shorthand
notation for a
$z$-derivative. Next, making the substitution
\be
\Omega = \frac{Q^{\prime}(z)}{Q(z)}
\ee
we obtain
\beq
Q^{\prime\prime} & = & 0 \label{eqeuler} \\
\frac{Q^{\prime}}{Q} + 2 \frac{A^{\prime}}{A} & = & 0 \label{eqar}
\eeq
Note that Eq.~(\ref{eqeuler}) is the Euler equation for ray
propagation in a homogeneous medium with general
solution $Q(z) = \alpha z + \beta $; we will be able
to interpret $Q(z)$ below as describing a ray nearby the periodic
orbit with relative angle and intercept
determined by $\alpha,\beta$.

\subsection{Boundary Conditions}
Having found the general solution of Eq.~(\ref{eqred2}) along one
segment of the periodic orbit we must impose
the boundary condition Eq.~(\ref{eqbc}) in order to connect solutions
in successive segments.
Writing out this condition:
\begin{displaymath}
\frac{c_m}{\sqrt{Q_m(z_m)}}\exp{\left(ikz_m+\frac{i}{2}\Omega(z_m)\tilde{x}_m^2\right)} +
\frac{c_{m+1}}{\sqrt{Q_{m+1}(z_{m+1})}}\exp{\left(ikz_{m+1}+\frac{i}{2}\Omega(z_{m+1})\tilde{x}_{m+1}^2\right)}|_{\partial D}= 0
\end{displaymath}
This equation must be satisfied on an arc of the boundary of length
$\sim \sqrt{\lambda l}$ around the reflection
point.  Since this length is much smaller than $l$ we can express
this arc on the boundary $\partial D$ as an
arc of a circle of radius $\rho$ (the curvature at the reflection
point).  We express the boundary condition
in a (scaled) common local
coordinate system for the incident and reflected fields
$(\tilde{\xi}_1,\tilde{\xi}_2)$ pointing along the tangent and the
normal at the bounce point (see Fig.~\ref{figcs}).  Because the boundary
condition must be satisfied on the entire arc
it follows the phases of each term must be equal,
\beq
\lefteqn{k\left(l_m + \frac{1}{\sqrt{k}}\tilde{\xi}_1\sin\chi_m -
\frac{1}{k}\frac{\tilde{\xi}_1^2}{2\rho_m}\cos\chi_m \right) +
\frac{1}{2}\frac{Q^{\prime}_m}{Q_m} \left(\tilde{\xi}_1 \cos\chi_m +
\frac{1}{k}\frac{\tilde{\xi}_1^2}{2\rho_m}\sin\chi_m\right)^2=}
\nonumber \\
& & k\left(l_m + \frac{1}{\sqrt{k}}\tilde{\xi}_1 \sin\chi_m +
\frac{1}{k}\frac{\tilde{\xi}_{1}^2}{2\rho_m}\cos\chi_m \right) +
\frac{1}{2}\frac{Q^{\prime}_{m+1}}{Q_{m+1}}\left(\tilde{\xi}_1
\cos\chi_m +
\frac{1}{k}\frac{\tilde{\xi}_1^2}{2\rho_m}\sin\chi_m\right)^2
\nonumber
\eeq
and there is a amplitude condition as well,
\begin{displaymath}
\frac{c_m}{\sqrt{Q_m(L_m)}} + \frac{c_{m+1}}{\sqrt{Q_{m+1}(L_{m})}} = 0
\end{displaymath}
Here, we assume that $\tilde{\xi}_1=O(1)$, and we will carry out the
solution of  these equations to $O(1/\sqrt{k})$.
Note that, it is sufficient to take $Q(z)|_{\partial D} \approx
Q(l_m)$, at this level.
In each segment we have three constants which determine our solution:
$c_m, \alpha_m, \beta_m$ (where $Q_m(z)=
\alpha_m z + \beta$); however due to its form our solution is
uniquely determined by the two ratios
$\beta_m/\alpha_m, c_m/\sqrt{\alpha_m}$.  Therefore we have the
freedom to fix one matching relation for the $Q_m$
by convention, which then determines the other two uniquely.  To
conform with standard definitions we fix $Q_{m+1}= Q_m$, then the
amplitude and phase equality conditions give the relations:

\be
\left(
\begin{array}{c}
Q_{m+1}\\
Q^{\prime}_{m+1}
\end{array}
\right) =
\left( \begin{array}{cc}
1 & 0 \\
-\frac{2}{\rho_m\cos\chi_m} & 1
\end{array}\right)
\left(
\begin{array}{c}
Q_{m}\\
Q^{\prime}_{m}
\end{array}
\right)
\equiv {\cal R}_m
\left(
\begin{array}{c}
Q_{m}\\
Q^{\prime}_{m}
\end{array}
\right)
\label{eqrABCD}
\ee
  and
\be
C_{m+1} = e^{-i\pi} C_m \label{eqtrans3}
\ee

Here we always choose the principal branch of $\sqrt{Q}$.  These
conditions allow us to propagate any solution through a reflection point
via the reflection matrix ${\cal R}_m$ and Eq. (\ref{eqtrans3}).
Note that with these conventions the reflection matrix is precisely the
standard ``ABCD'' matrix of ray optics for reflection at arbitrary
incidence angle (in the tangent plane) from a curved mirror
\cite{siegman_book}.  It
follows  that $Q_m$ and $Q_{m+1}$ can be interpreted as the
transverse coordinates
of the incident and reflected rays with respect to the PO, and that
then Eq.~(\ref{eqrABCD}) describes specular
ray reflection at the boundary, if the non-linear dynamics is linearized around the reflection point.

\subsection{Ray dynamics in phase space}

To formalize the relation to ray propagation in the paraxial limit we
define the ray position coordinate
to be $Q$ and the conjugate momentum $P=Q'$, with $z$
playing the role of the time.
Let's introduce the
column-vector $(\stackrel{\scriptstyle \:Q\:}{_{\scriptstyle
\:P\:}})$. The ``fundamental matrix" $\Pi$ is the
matrix obtained by two such linearly independent vectors
\be
\Pi = \left(
\begin{array}{cc}
Q_1 & Q_2 \\
P_1 & P_2
\end{array}
\right)
\ee
where the linear independence, as expressed by the Wronskian
condition $W(Q^{\prime}, Q) \neq 0$ reduces to $\mbox{det}\Pi \neq
0$. Then, for example, the Euler
equation  Eq.~(\ref{eqeuler}) in each arm can be expressed as
\be
\frac{d\Pi}{dz} = {\cal JH}\,\Pi
\label{eqmotion}
\ee
where ${\cal H}=(\:\stackrel{\scriptstyle 0\: 1}{_{\scriptstyle 0\:
0}}\:)$ and ${\cal J} = (\:\stackrel{\scriptstyle 1\mbox{\ }
0}{_{\scriptstyle 0 -\!1}}\:)$. It's straightforward to show that
$\mbox{det}\Pi$ is a constant of motion. To
take into account the discreteness of the dynamics in a natural
manner, we will introduce ``coordinates" for the ray, represented by
a $z$-independent column vector $h$.
Any ray of the $m$th arm in the solution space of Eq.~(\ref{eqeuler})
can then be expressed by a z-independent column vector $h_m$
\be
\left(
\begin{array}{c}
Q(z)\\
P(z)
\end{array}
\right) = \Pi(z) h_m
\ee
     We will choose $\Pi(z)=(\:\stackrel{\scriptstyle 1\:
z}{_{\scriptstyle 0\: 1}}\:)$, so that $\Pi^{-1}(z) =
\Pi(-z)$. Then if
$Q_m(z)=\alpha_m z +\beta_m$, we have
$h_m = (\stackrel{\scriptstyle
\:\beta_m\:}{_{\scriptstyle
\:\alpha_m\:}})$. In this notation, propagation within each arm (i.e.
$z+l < L_m$ where $L_{m-1}<z<L_m$) is induced by the `propagator' $\Pi$
\be
\left(
\begin{array}{c}
Q_m(z+l)\\
P_m(z+l)
\end{array}
\right) = \Pi(l) \left(
\begin{array}{c}
Q_m(z)\\
P_m(z)
\end{array}
\right)
\ee
which is the ``ABCD'' matrix for free propagation \cite{siegman_book}.
Thus, for example
$z\rightarrow z^{\prime}$ propagation for $L_{m-1}<z<L_m<z^{\prime}<L_{m+1}$ is
\be
\left(
\begin{array}{c}
Q_{m+1}(z^{\prime})\\
P_{m+1}(z^{\prime})
\end{array}
\right) = \Pi(z^{\prime}-L_m) {\cal R}_m \Pi(L_m-z) \left(
\begin{array}{c}
Q_m(z)\\
P_m(z)
\end{array}
\right)
\ee
Of special importance is the monodromy matrix, ${\cal M}(z)$, which
propagates rays a full round-trip,
i.e. by the length $L$ of the corresponding PO and is given by
\be
{\cal M}(z) = \Pi(z-L_{m-1}){\cal R}_{m-1}
\Pi(l_{m-1})\cdots\Pi(l_{m+1}){\cal R}_m\Pi(L_m-z)
\ee
for $L_{m-1}<z<L_m$. Note that $l_{m+N}=l_m$ and ${\cal
R}_{m+N}={\cal R}_m$. Although the specific
form of ${\cal M}(z)$ depends on the choice of origin, $z$, it is
easily shown that all other
choices would give a similar matrix and hence the eigenvalues of the
monodromy matrix are independent of this choice.  We will
suppress the argument $z$ below.

\subsection{Single-valuedness and quantization}

\label{section:quant1}
Having determined how to propagate an initial solution $Q,P$ an
arbitrary distance around the
periodic orbit we can generate a solution of the parabolic equation
which satisfies the
boundary conditions by an arbitrary initial choice $Q(0),P(0)$.
However an arbitrary
solution of this type will not reproduce itself after propagation by
L (one loop around the PO).
Recalling that our solution is translated back into the
two-dimensional space $(X,Z)$ by
Eq.~(\ref{eqans1}), and that the function $E(X,Z)$ must be single-valued, we
must require periodicity for
our solutions
\be
E(x,z+L)=E(x,z).
\label{eqper}
\ee
We will suppress again the reference to arm index and use the
notation $E(x,z)=E_m(x_m,z_m)$ and $u(x,z)=u_m(x_m,z_m)$ whenever
$L_{m-1} < z < L_m$.
Since the phase factor in $E$ advances by $\exp[ikL]$ with each
loop around the periodic orbit and 
single-valuedness implies that
\be
u(x,z + L)e^{ikL} = u(x,z).
\ee
This periodicity condition will only be solvable for
discrete values of $k$ and will lead to our quantization rule for the PO modes.

    From the form of $u$ in Eq.~(\ref{eqans3}) we see that the phase
$\Omega (z) =
Q^{\prime}/Q$ will be unchanged
if we choose $(Q,P)$ to be an eigenvector $(q_1,p_1)),(q_2,p_2)$ of
the monodromy matrix as in this
case the $Q^{\prime}(z+L) = \lambda_{1,2}Q^{\prime}(z),Q(z+L) =
\lambda_{1,2}Q(z)$
and the ratio $\Omega (z) =
\Omega (z+L)$.  The monodromy matrix is unimodular and symplectic and
its eigenvalues come
in inverse pairs, which are either purely imaginary (stable case) or
purely real (unstable
and marginally stable cases).  If the PO is unstable and the
eigenvalues are real, then the
eigenvectors and hence $\Omega (z)$ are real.  But a purely real
$\Omega (z)$ means that the
gaussian factor in $u(x,z)$ is purely imaginary and the solution does
not decay in the direction transverse
to the PO, contradicting the initial assumption of the parabolic
approximation to the Helmholtz
equation.  {\it Hence for unstable POs our construction is inconsistent
and we cannot find a solution of this form localized near the PO.}

On the other hand, for the stable case the eigenvalues
come in complex conjugate
pairs and the reality of the monodromy matrix then implies that the
{\it eigenvectors} cannot
be purely real and are related by complex conjugation.
Therefore we can always construct a gaussian solution
based on the eigenvector with
$Im{\Omega} > 0$ so that  there is a negative
real part of the gaussian exponent and the resulting solution decays
away from the PO.    It is
easy to check that the range of the decay is $\sim \sqrt{\lambda l}$
as assumed above.

The eigenvectors of the monodromy matrix are sometimes referred to as
Floquet rays of the
corresponding Euler equation.  In a more formal discussion which can
be straightforwardly generalized
to arbitrary index variation and three dimensions
\cite{babic_book} our condition on the choice of $\Omega$
can be expressed by recalling that the two eigenvectors must generate
a constant Wronskian,
and we will choose the eigenvector $(q(z),p(z))$ such that
\be
pq^*-qp^*=i.
\label{eqcom}
\ee
This condition implies $Im{\Omega} > 0$.
Note that the overall magnitude of the Wronskian is determined by our
choice of normalization
of the eigenvectors and is here chosen to be unity.

Having made this uniquely determined choice (up to a scale factor) and
using $(\stackrel{\scriptstyle \:q(z+L)\:}{_{\scriptstyle \:p(z+L)\:}})
=e^{i\varphi}(\stackrel{\scriptstyle \:q(z)\:}{_{\scriptstyle
\:p(z)\:}})$ and $c_{m+N}=e^{-i\pi N }c_m$
we obtain
\be
u(\tilde{x},z+L) = e^{-i\frac{\varphi}{2} - i\pi (N_{\mu} + N)}u(\tilde{x},z)
\ee

   From Eq.~(\ref{eqper}) and Eq.~(\ref{eqans2}), we finally obtain the
quantization rule for the wavevectors
of the bound states of the closed cavity:
\be
kL=\frac{1}{2}\varphi + 2\pi m + \bmod_{2 \pi}[(N + N_{\mu})\pi]
\label{quant1}
\ee
where $m$ is an integer, $N$ is the number of bounces in the PO, $\varphi$ is
the Floquet phase obtained from the eigenvalues of the monodromy matrix and
$N_{\mu}$ is an integer known as the Maslov index in the non-linear dynamics
literature \cite{maslov_book}.  It arises in the following manner.
As already noted, the
eigenvalues of the monodromy matrix for stable POs are complex conjugate
numbers of modulus unity whose phase is the Floquet phase.  We can
define $\varphi =
\mbox{Arg}[\sqrt{q(L)/q(0)}]$, where $\mbox{Arg}[\cdot]$ denotes the principal
argument; hence the Floquet phase depends only on $\cal{M}$.
However our solution, Eq.~(\ref{eqans3}), involves the {\it square root}
of $q(z)$ and will be sensitive to
the number of times the phase of $q(z)$ wraps around the origin as $z$ goes
from zero to $L$.  If this winding number (or Maslov index) is called
$N_{\mu}$ then
the actual phase advance along the PO is $\varphi + 2 \pi N_{\mu}$; if
$N_{\mu}$ is odd this leads to an observable $\pi$ phase shift in the
solution not
included by simply diagonalizing ${\cal M}$ to find the Floquet phase.

$N_{\mu}$ may be directly calculated by propagating $q(z)$:
\be
  N_{\mu} = \left[\frac{1}{2\pi i} \int_0^L d(\ln{q(z)})\right]
\ee
where $[\cdot]$ denotes the integer part. There is no simple rule for
reading off the Maslov index from the geometry of the PO; however the
Maslov index doesn't affect the free spectral range or the transverse
mode spacing.

\begin{figure}[h]
\centering
\includegraphics[width=\linewidth]{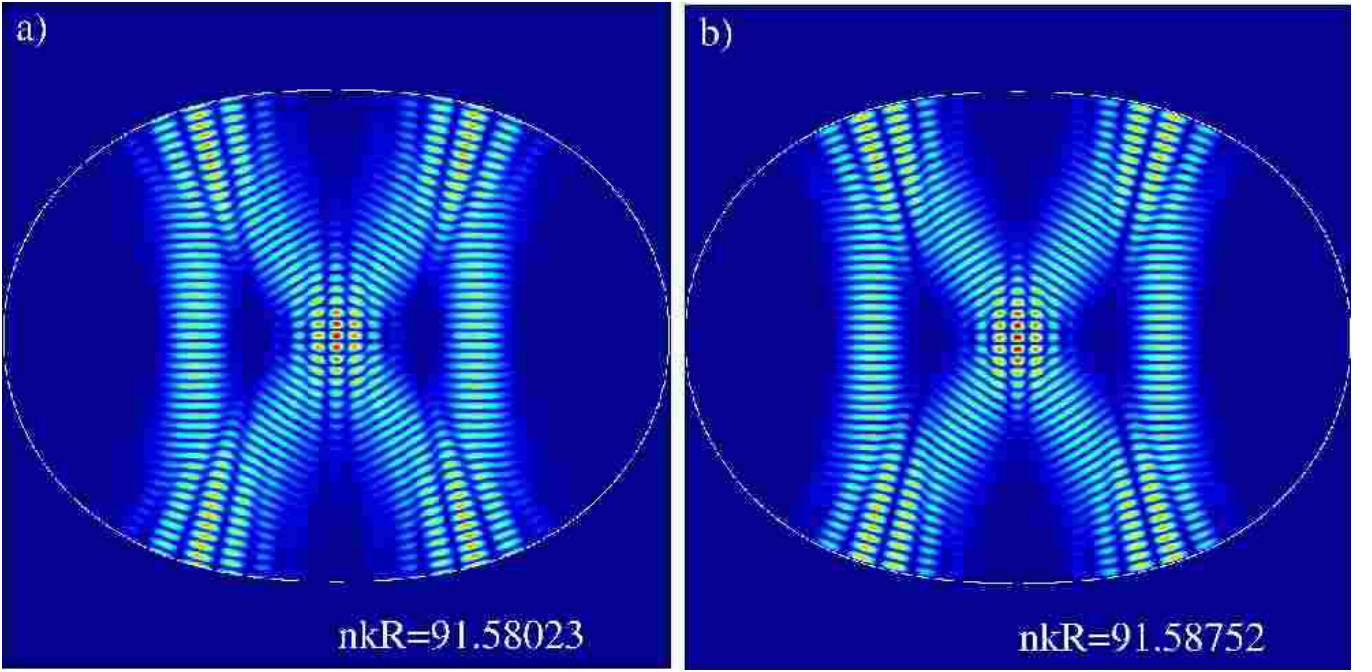}
\caption{Intensity of the TM solution for a bow-tie mode plotted
in a false color scale, (a) calculated numerically and (b) from the gaussian
optical theory with parameters $m=100$, $\varphi=2.11391$, $N_{\mu}=1$ and $N=4$. Note the excellent agreement of the quantized values
for $kR$ (R is the average radius of the quadrupole).}
\label{fig:compg}
\end{figure}

Except for the Maslov index, the quantization rule Eq.~(\ref{quant1})
is familiar
from Fabry-Perot resonators.  The longitudinal mode index
$m$ gives rise to a free spectral range
$\Delta k_{long} = 2 \pi/L$ for gaussian modes of a stable PO of length $L$.
The Floquet phase $\varphi/2L$ is
the zero-point energy associated with the transverse quantization of
the mode and we will shortly derive excited transverse modes with spacing
$\varphi/L$.  In Fig.~\ref{fig:compg}, we plot for comparison the analytic solutions
for the bow-tie resonance just derived in comparison to a numerical
solution of the same problem; both the intensity patterns and the
quantized value of $k$ agree extremely well.

\subsection{Transverse excited modes}
As is well known, the solution Eq.~(\ref{eqans3}) is only one family
of the possible solutions of the parabolic
equation  Eq.~(\ref{eqred2}) satisfying the boundary conditions
Eq.~(\ref{eqbc}) and the periodicity condition
Eq.~(\ref{eqper}), the one corresponding to the ground state for
transverse oscillations. Further solutions can be
generated with algebraic techniques which were originally developed
in the context of quantum oscillators \cite{chernikov1}.
If we refer to Eq.~(\ref{eqred2}) we see that we are looking for solutions of
${\cal L} (u) = 0$, i.e. eigenfunctions
of the differential operator with eigenvalue zero.  It is natural
following the analogy to
quantum oscillators to seek additional solutions by defining lowering
and raising operators.
\be
\Lambda(z) = -iq(z)\partial_{\tilde x} + p(z)\tilde{x}
\label{eqannh}
\ee
\be
\Lambda^{\dagger}(z) = iq^*(z)\partial_{\tilde x} + p^*(z)\tilde{x}.
\label{eqannhd}
\ee
We can easily show that the operators
($\Lambda,\Lambda^{\dagger},{\cal L}$) form an algebra. Namely,
that $[\Lambda^{\dagger},{\cal L}] = [\Lambda,{\cal L}]=0$ and furthermore that
$[\Lambda,\Lambda^{\dagger}]=i(p^*q-q^*p)=1$.
Defining the ``ground-state'' solution we have found as $u^{(0)}$, the
commutation condition implies that
$(\Lambda^{\dagger} u^{(0)}) (x,z)$ is also a solution of ${\cal L} (u) =
0$. Further it can be
checked that while $(\Lambda u^{(0)})  = 0, (\Lambda^{\dagger}) u^{(0)} $ is a
non-trivial solution.
We can find the wavevector quantization condition for this solution
by calculating the
additional phase acquired by $(\Lambda^{\dagger} u^{(0)}) (x,z)$ upon
performing one loop around
the orbit $(\Lambda^{\dagger} u^{(0)}) (x,z+L)$.  Noting that $q^*(z+L) =
e^{-i\varphi} q^*(z),p^*(z+L) =
e^{-i\varphi} p^*(z)$, we find that $ \Lambda^{\dagger} (z+L) =
e^{-i\varphi} \Lambda^{\dagger}(z)$.
Thus this solution will acquire an additional phase $-\varphi$ with
respect to $u^{(0)}$.  This means that the $l^{th}$ family
of solutions
\be
u^{(l)}(\tilde{x},z)=(\Lambda^{\dagger})^l u^{(0)}
\ee
will satisfy the wavevector quantization rule:

\be
kL= (l+\frac{1}{2}) \varphi + 2\pi m + \bmod_{2 \pi}[(N + N_{\mu})\pi]
\label{quant2}
\ee

This result has the well-known interpretation of adding $l$
transverse quanta $k_T=\varphi/L$ to the energy of the
ground state gaussian solution.  Thus (in two dimensions)
the general gaussian modes have two modes indices $(l,m)$
corresponding to the number of transverse and longitudinal quanta
respectively and two different
uniform spacings in the spectra (see Fig.~\ref{fig:spectrum}(b)).

In order to obtain explicit forms for the excited state solutions found here
one may use the theory of orthogonal polynomials to find
\be
u^{(l)}(\tilde{x},z)=\left(i\sqrt{\frac{q^*(z)}{2q(z)}}\;\right)^l
H_l\left(\sqrt{Im\left[\frac{p(z)}{q(z)}\right]}
     \tilde{x} \:\right)\:u^{(0)} (\tilde{x},z)
\label{eqexc}
\ee
where $H_l$ are the Hermite polynomials.

As noted in the introduction, the solutions we have found by the parabolic
equation method do not reflect correctly the discrete symmetries
of the cavity which may be present.  The theory of the symmetrized
solutions and the breaking of degeneracy is essentially the same
for both the closed and open cavity and will be presented in
section (\ref{section:sym1}) after treating the open case in the next section.

\section{Opening the cavity - The dielectric resonator}

We now consider the wave equation for a uniform dielectric rod of
arbitrary cross-section surrounded by vacuum.
As noted in the introduction, Maxwell's equations separate into TM and TE
polarized waves satisfying
\begin{equation}
\left( \nabla^2 + n({\bf r})^2k^2 \right) \Psi = 0
\label{hh}
\end{equation}
Maxwell boundary conditions translate into continuity of $\Psi$
and its normal derivative across the boundary $\partial D$, defined
by the discontinuity in $n$. Here
$\Psi=E$($\Psi=B$) for $TM$($TE$) modes.  We will only consider the
case of a uniform dielectric
in vacuum for which the index of refraction $n ({\bf r}\in D) = n$
and $n ({\bf r}\not\in D) = 1$.  Thus we have the Helmholtz equation
with wave vector $nk$ inside the dielectric
and $k$ outside.  The solutions to the wave equation for this case
cannot exist only within the dielectric,
as the continuity conditions at the dielectric interface do not allow
such solutions.  On physical
grounds we expect solutions at every value of the external wavevector
$k$, corresponding to elastic
scattering from the dielectric, but that there will be narrow intervals $\delta
k$ for which these solutions will have
relatively high intensity within the dielectric, corresponding to the
scattering resonances.  A
standard technique for describing these resonances as they enter into
laser theory is to impose the boundary
condition that there exist only outgoing waves
external to the cavity.  This boundary
condition combined with the continuity conditions cannot be satisfied
for real wavevectors $k$ and
instead leads to discrete solutions at complex values of $k$, with
the imaginary part of $k$ giving
the width of the resonance.  These discrete solutions are called
quasi-bound modes or quasi-normal modes \cite{young1}.
We now show how such quasi-bound modes can be incorporated into the
gaussian optical resonance theory just described.

As before we look for solutions which, within the cavity, are
localized around periodic orbits based
on specular reflection within the cavity.  In order to satisfy the
boundary conditions the solutions
outside the cavity will be localized around rays extending out to
infinity in the directions determined
by Snell's law at the bounce point of the PO.  The quasi-bound state
condition is imposed by insisting
that the solutions along those rays to infinity are only outgoing.
This can only be achieved by
making $k$ complex (assuming real index $n$).  It is worth noting
that there are well-known corrections
to specular reflection and refraction at a dielectric interface, for
example the Goos-Hanchen shift \cite{ra1}.
It is interesting to attempt to incorporate such effects into our
approach at higher order, however we
do not do so here and confine ourselves to obtaining a consistent solution
to lowest order in $kl$.

As before we will define the solution as the sum of solutions
$E_m(x_m,z_m)$ attached to each segment of the
PO and define local coordinates $(x_m,z_m)$ attached to the PO.   Now
in addition we need an outside solution
at each reflection point $E_{mt}$ with its own coordinate system
$(x_{mt},z_{mt})$ rotated by an angle given by
Snell's law applied to the direction of the incident ray (see
Fig.~\ref{figcs}). The ansatz of
Eq.~(\ref{eqans2}) thus applies, where now the sum will run over
2N components, which will include the
transmitted fields. Introducing the slowly varying envelope
approximation Eq.~(\ref{eqans1}) and the scalings
$\tilde{x}_{mt}=\sqrt{k}x$, $\tilde{x}_{m}=\sqrt{nk}x$, we get the
parabolic equation Eq.~(\ref{eqred2}) for each
component, at lowest order in $k$. The boundary conditions close
to the $m^{\mbox{th}}$ bounce point will take
the form
\be
E_i + E_r |_{\partial D^-} = E_t |_{\partial D^+}
\label{eqbc1}
\ee
and
\be
\partial_n E_i + \partial_n E_r |_{\partial D^-} = \partial_n E_t
|_{\partial D^+}
\label{eqbc2}
\ee
Here $\partial_n$ is the normal derivative at the boundary.
The alternative indices $i$, $r$, $t$ stand for $m$, $m+1$ and $mt$,
respectively. Since the parabolic equation
Eq.~(\ref{eqred2}) is satisfied in appropriately scaled coordinates
within each segment, we write all solutions in
the general form $E_M=A_M\;exp{(i\Phi_M)}$ where $\Phi_M = n_Mkz +
\frac{i}{2}Q^{\prime}_M Q_M^{-1}\tilde{x}^2_M$
and $A_M = c_m/\sqrt{Q_M}$. Here $M$ stands for $m$ or $mt$ and
$n_m=n$, $n_{mt}=1$. As for the closed case,
we need to
determine $Q_M, Q^{\prime}_M$ and $c_M$, so that the boundary conditions are
satisfied, and then impose single-valuedness to quantize $k$.

Similarly to the closed case, the first continuity condition Eq.
(\ref{eqbc1}) must be satisfied on an arc
of size $\sim \sqrt{{\lambda} l}$ on the boundary around each bounce
point and that implies that the phases
of the incident, transmitted and reflected waves must be equal.  This
equality will be implemented in
the coordinate system  $(\tilde{\xi}_1,\tilde{\xi}_2)$ along the
tangent and normal to the boundary
at the reflection point, as before.  We can again expand the boundary
as the arc of a circle of radius
$\rho$, the curvature at the bounce point.  Since the equations are
of the same form for each reflection
point it is convenient at this point to suppress the index $m$ and
use the indices $i,r,t$ to denote
the quantities associated with the incident, reflected and
transmitted wave at the $m^{th}$ bounce point.

The analysis of the phase equality on the boundary for the incident
and reflected waves is exactly the
same as in the closed case and again leads to Eq.~(\ref{eqrABCD}) describing specular
reflection. Equating the phases of the incident and transmitted wave leads to:
\beq
\Phi_i & = & nk\left(L_m + \frac{1}{\sqrt{nk}}\tilde{\xi}_1\sin\chi
- \frac{1}{nk}\frac{\tilde{\xi}_1^2}{2\rho}\cos\chi \right) +
\frac{1}{2}\frac{Q^{\prime}_i}{Q_i} \left(\tilde{\xi}_1 \cos\chi +
\frac{1}{nk}\frac{\tilde{\xi}_1^2}{2\rho}\sin\chi\right)^2
\nonumber \\
\Phi_t & = & k\left(nL_m + \frac{1}{\sqrt{nk}}\tilde{\xi}_1\sin\chi_{t} -
\frac{1}{nk}\frac{\tilde{\xi}_1^2}{2\rho}\cos\chi_{t} \right) +
\frac{1}{2}\frac{Q^{\prime}_{t}}{Q_{t}}
\left(\tilde{\xi}_1 \cos\chi_{t} +
\frac{1}{nk}\frac{\tilde{\xi}_1^2}{2\rho}\sin\chi_{t}\right)^2
\nonumber
\eeq
where we have defined $\chi=\chi_m, \rho=\rho_m$ and
all quantities are evaluated at $z=L_m$.
Recalling that $n\sin\chi_i = \sin\chi_{t}$ we get up to  $O(1/\sqrt{k})$
\be
\left(
\begin{array}{c}
Q_t\\
Q^{\prime}_t
\end{array}
\right) =
\left( \begin{array}{cc}
\ 1/\mu & 0 \\
-\frac{2(1-\mu)}{\rho\cos\chi} & n\mu
\end{array}\right)
\left(
\begin{array}{c}
Q_i\\
Q^{\prime}_i
\end{array}
\right)
\label{eqtABCD}
\ee
where $\mu = \cos\chi_i / \cos\chi_{t}$ and the relation $ Q_i  = 
\mu Q_{t}$ is a
convention similar to $Q_r=Q_i$.
Again, the  matrix in Eq.~(\ref{eqtABCD}) is just the ABCD matrix for 
transmission of rays through
a curved dielectric interface at arbitrary angle of incidence in the
tangential plane \cite{siegman_book}.

Using the phase equality on the boundary the continuity of the field
gives the general transport equation:
\be
\frac{c_i}{\sqrt{Q_i}} +\frac{c_{r}}{\sqrt{Q_{r}}} =
\frac{c_{t}}{\sqrt{Q_{t}}}
\ee
which becomes (using the conventions $Q_r = Q_i,Q_t = Q_i/\mu$)
\be
c_i + c_{r} = \sqrt{\mu}c_{t}
\label{eqtr1}
\ee
In order to find the quantization condition we need a direct relation
between $c_i$ and $c_r$ as we had in the
closed case.  This is provided by the normal derivative boundary
condition Eq.~(\ref{eqbc2}). Keeping only the leading terms this condition
becomes:
\be
c_i \partial_n \Phi_i + c_r \partial_n \Phi_r = \sqrt{\mu}
c_t \partial_n \Phi_t.
\ee
At the level of approximation needed one finds the simple results $
\partial_n \Phi_i = nk \cos \chi,
     \partial_n \Phi_r = -nk \cos \chi,  \partial_n \Phi_t = k \cos
\chi_t$, leading to
\be
n\sqrt{\mu}(c_i-c_{r}) = c_{t}
\label{eqtr2}
\ee
and
\be
c_r = \frac{n\mu-1}{n\mu+1} c_i.
\label{eqfres}
\ee
Note the key result that
\be
|c_r|^2 = \frac{|n \cos \chi_i - \cos \chi_t|^2}{|n \cos \chi_i + \cos
\chi_t|^2} |c_i|^2
\ee
which is precisely the Fresnel reflection law at a (flat) dielectric interface.

Now we impose the single-valuedness or periodicity condition to obtain
the quantization rule for
$k$.
\be
E(x,z+L) = E(x,z)
\ee
Note however that we have a qualitatively different situation than
for the closed cavity; some amplitude can
be lost at each reflection and it will in general be impossible to
make a loop around the PO and return to the
same field amplitude unless $k$ is complex.  We have the condition
\be
u(x,z+L)e^{inkL} =  u(x,z)
\ee
which can be satisfied by choosing $Q(z),P(z)$ to be the appropriate
eigenvector of the monodromy matrix
(note that ${\cal M}$ is unchanged from the closed case as it only pertains to
the propagation of the phase) and
with this choice the quantization condition becomes
\be
nkL=\frac{1}{2}\varphi  + 2\pi m  +  \bmod_{2 \pi}[(N + N_{\mu})\pi]
-i\sum_{b=1}^N \log{\left[\frac{n\mu_b-1}{n\mu_b+1}\right]}.
\label{quant3}
\ee

Recall that the Fresnel reflection law has the property that it gives
a pure phase for rays incident
above total internal reflection.  Thus the new term in the
quantization law due to Fresnel reflection
can be either purely real (all bounces of PO totally internally
reflected), purely positive imaginary
(all bounces below TIR) or complex (some TIR bounces, some refracted
bounces).  If we define:
$ \varphi_f = \mbox{Re}[-i\sum_b^N
\log{\left[\frac{n\mu_b-1}{n\mu_b+1}\right]}] $ and
$ \gamma_f = \mbox{Im}[-i\sum_b^N
\log{\left[\frac{n\mu_b-1}{n\mu_b+1}\right]}] $ then the quantization
rule gives
\beq
\mbox{Re} [nkL]  & = & 2\pi m + \bmod_{2 \pi}[(N + N_{\mu})\pi]  + \varphi/2 +
\varphi_f \\
\mbox{Im} [nkL] & = & -\gamma_f.
\eeq

As noted above, this result is only in the leading order
approximation, and it ignores both the effects of evanescent leakage
at a curved interface and the momentum width of the gaussian
``beam'' which leads to violations of ray optics.  These
effects will give a non-zero imaginary part (width) to all 
resonances, even those with
all bounce points above TIR.

In table ~\ref{tableopen} we present a comparison between the 
numerically obtained quantized
wavevectors for a bow-tie resonance and the values for the real and 
imaginary part of $k$
predicted by Eq.~(\ref{quant3}) for three different indices of 
refraction.  Note
that the best agreement is for the case far from total internal reflection,
and the worst agreement is the case near TIR. 

\begin{table*}[h]
\begin{center}
\begin{tabular}{|c|c|c|c|}\hline
index&numerical&gaussian&surface of\\
of refraction & calculation & quantization rule &section\\ \hline
\rule[-4mm]{0mm}{10mm}$n=2.0$&$nkR=100.53788+0.49758i$&$nkR=100.53858+0.49635i$&\\ \cline{1-3}
\rule[-4mm]{0mm}{10mm}$n=2.9$&$nkR=100.59376+0.16185i$&$nkR=100.53858+0.14178i$&\\ \cline{1-3}
\rule[-4mm]{0mm}{10mm}$n=5.1$&$nkR=100.84716+0.00245i$&$nkR=100.85111+0.00000i$&
\raisebox{-2ex}[-1.5ex]{\includegraphics[width=2.5cm,height=2.8cm]{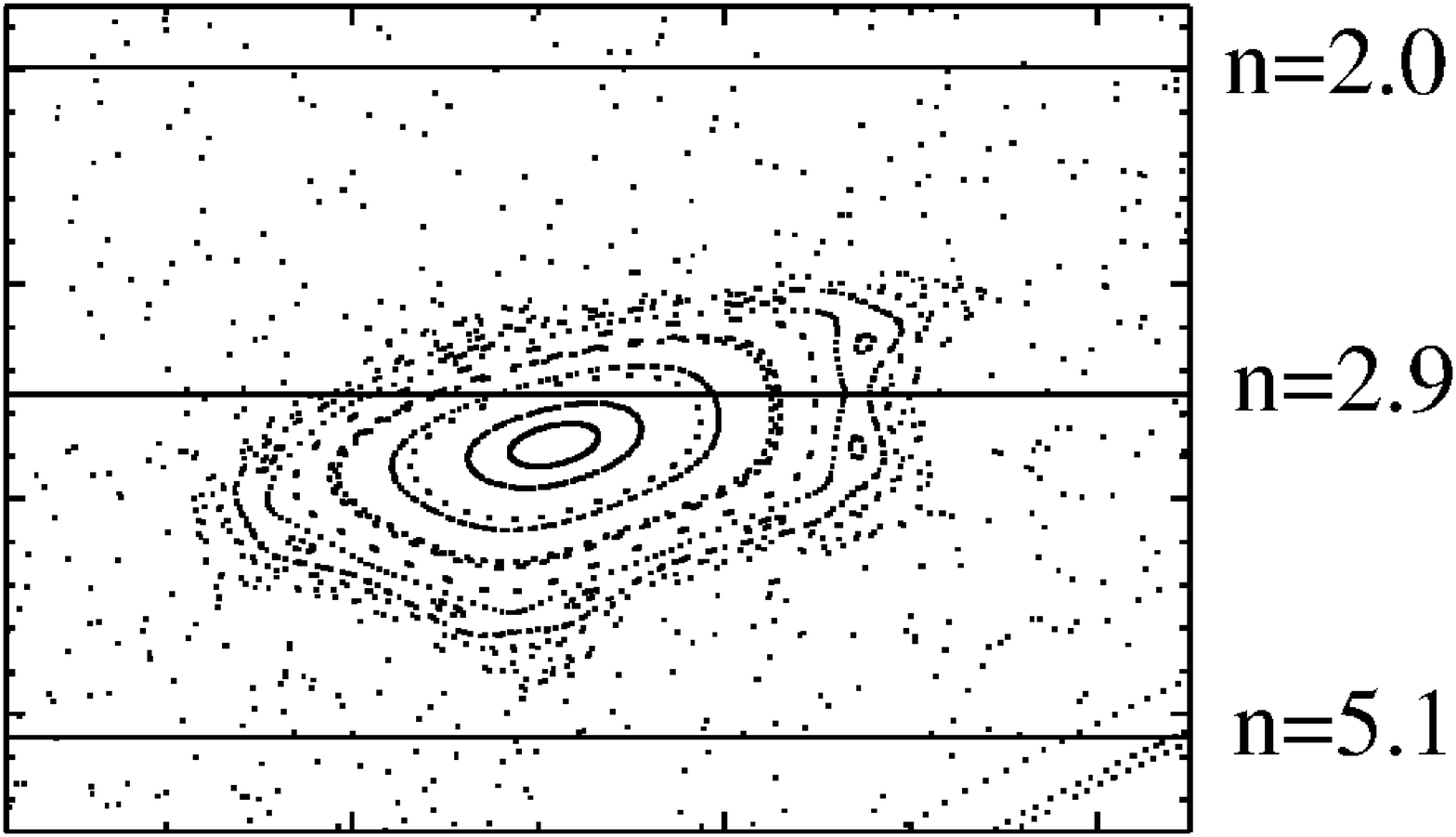}}\\ \hline
\end{tabular}
\end{center}
\caption{Table comparing the gaussian optical prediction for the
complex $k$ values of bow-tie resonances of the quadrupole ($\epsilon = 0.17$) to
numerically obtained values for three different indices of refraction 
corresponding
to incidence below, at and above the critical angle for total 
internal reflection.
The schematic at right depicts the bow-tie island in comparison to the critical
line for total internal reflection for the three cases (horizontal lines).}
\label{tableopen}
\end{table*}

\section{Symmetry Analysis and Quasi-Degeneracy}
\label{section:sym1}
As noted above, the gaussian theory we have just presented predicts
exact degeneracies if there exist
several symmetry-related stable orbits (note that in this context the
same path traversed
in the opposite sense is
considered a distinct symmetry-related orbit).  However it is well-known that
wave equations with discrete symmetries cannot have a degeneracy
which is larger than the largest
dimension of the irreducible representations of the symmetry group of
the equation.  Here we are
concerned with the point group (rotations and reflections) of the
dielectric resonator in two dimensions.
Let G be a group which leaves the cavity invariant. Then, if
$\Psi(\bf{x})$ is a solution to the
Helmholtz equation, so is $\Psi(g\bf{x})$, where
$g\in G$ and ${\bf x}=(X,Z)$. The symmetrized
solutions, i.e. solutions which transform according to the
irreducible representations of $G$ under the action of
$G$, can be obtained by the projection operators of the group:
\be
{\cal P}_m E({\bf x}) = \frac{d_m}{|G|} \sum_{g\in G} \chi_m(g) E(g{\bf x})
\label{eqproj}
\ee
Here $d_m$ is the dimensionality and $\chi_m(g)$
is the character of the m$^{th}$ irreducible representation and the
solution so obtained, denoted by $E_m({\bf x})$
is the resulting symmetry-projected solution. For a given irreducible
representation there are as many
symmetrized solutions as the dimension of that irreducible
representation of $G$. We will focus here on the case of the closed resonator, but the general principles apply to the open case as well.

\subsection{Symmetrized modes for the quadrupole}
Let's consider our canonical example, the quadrupole (see Fig.~\ref{figcs}).
The symmetry group of the quadrupole
is $G = C_2 \otimes C_2=\{1,\sigma_X,\sigma_Z,\sigma_X\sigma_Z\}$,
the group of reflections about the $X$ and $Z$
axes. This group has four one-dimensional representations only, and
thus cannot have {\it any} exactly
degenerate solutions (barring accidental degeneracy).  The existence
of four irreducible representations means that
given the one solution $E(X,Z)$ we have constructed to the Helmholtz
equation, we can generate four linearly
independent solutions by projection according to Eq.~(\ref{eqproj})
above. We will
label the representations by $m=(r,s)$, where $r,s=\pm$ denotes the
action of inversion of $X$ and $Z$
respectively.  The symmetrized solutions are then
\be
\begin{array}{ll}
E_{(++)} = \frac{1}{4}(e_1 + e_2 + e_3 + e_4), &
E_{(+-)} = \frac{1}{4}(e_1 + e_2 - e_3 - e_4) \\
E_{(-+)} = \frac{1}{4}(e_1 - e_2 + e_3 - e_4), &
E_{(--)} = \frac{1}{4}(e_1 - e_2 - e_3 + e_4) \\
\end{array}
\label{projwf}
\ee
where
\be
\begin{array}{llll}
e_1 = E(X,Z), & e_2 = E(-X,Z), & e_3 = E(X,-Z), & e_4 = E(-X,-Z).
\end{array}
\ee
In this case the symmetrized solutions are just the solutions with
definite parity with
respect to the symmetry axes of the quadrupole.

The key point here is that the within the parabolic equation
approximation these four solutions are
exactly degenerate, whereas our group-theoretic analysis for the
exact Helmholtz equation tells us that they cannot be so, although
they will be nearly degenerate.  Moreover our original solution
$E(X,Z)$ cannot be an exact solution,
as it does not transform as {\it any} irreducible representation of
the symmetry group.
A further important point is that while we can always construct a
number of symmetrized
solutions equal to the sum of the dimensions of the irreducible
representations, there is no
guarantee that such a projection will yield a non-trivial solution.
In fact in the case of the
quadrupole we will show below that for each quantized value of $k$
only two of the projected solutions
are non-trivial, leading to quasi-degenerate doublets in the
spectrum.  We will present below a simple
rule which allows one to calculate the quasi-degeneracy given the
periodic orbit and the symmetry
group of the  resonator.

Before discussing the general rule, we illustrate the basic procedure
for the case of a bow-tie PO.
Let $\ell_1$ and $\ell_2$ be the lengths of the vertical and diagonal
legs of the bow-tie, so that $L=2(\ell_1 +
\ell_2)$ is the total length. Then,
\begin{itemize}
\item $g=1$
\be
e_1 = E(g {\bf x}) = \frac{1}{\sqrt{q(z)}}\exp{\left[ikz +
\frac{i}{2}\Omega(z)x^2\right]}
\ee
\item $g=\sigma_X$ : $(z\rightarrow L/2 + z, x\rightarrow -x)$
\be
e_2 = E(g {\bf x}) = e^{\frac{1}{2}i\pi}
\frac{1}{\sqrt{e^{i\varphi/2}q(z)}}\exp{\left[ik(z+\frac{L}{2} +
\frac{i}{2}\Omega(z)x^2\right]} \equiv e^{i\zeta} e_1
\ee
\item $g=\sigma_Z$ : $z\rightarrow \ell_1 - z, x\rightarrow x)$
\be
e_3 = E(g {\bf x}) = \frac{1}{\sqrt{q(\ell_1 -
z)}}\exp{\left[ik(\ell_1 - z) + \frac{i}{2}\Omega(\ell_1 -
z)x^2\right]}
\ee
\item $g = \sigma_X\sigma_Z$ : $(z\rightarrow L/2 + \ell_1 - z,
x\rightarrow -x)$
\be
e_4 = E(g {\bf x}) = e^{\frac{1}{2}i\pi}
\frac{1}{\sqrt{e^{i\varphi/2}q(\ell_1 -
z)}}\exp{\left[ik(\frac{L}{2} + \ell_1 - z) +
\frac{i}{2}\Omega(\ell_1 - z)x^2\right]} \equiv e^{i\zeta} e_3
\ee
\end{itemize}
where the phase factor $\zeta = 1/2\,(\pi - \varphi/2 + kL)$. Here we
use the fact that ${\cal M}_L = {\cal
M}_{L/2}{\cal M}_{L/2}$ for the bow-tie orbit, where $M_L$ is the
monodromy matrix for the whole length $L$. It
follows that
$q(z+L/2) = e^{i\varphi/2}q(z)$.
    Note also the appearance of the factors $e^{\frac{1}{2}i\pi}$, which
is due to the
specific choice of branch-cut for $\sqrt{q(z)}$. Putting these
together, we obtain
\be
\begin{array}{lcl}
E_{(++)} & = & \frac{1}{2}(e_1 + e_3) e^{i\frac{\zeta}{2}} \cos
\frac{\zeta}{2} \\
E_{(+-)} & = & \frac{1}{2}(e_1 - e_3) e^{i\frac{\zeta}{2}} \cos
\frac{\zeta}{2} \\
E_{(-+)} & = & \frac{1}{2i}(e_1 + e_3) e^{i\frac{\zeta}{2}} \sin
\frac{\zeta}{2} \\
E_{(--)} & = & \frac{1}{2i}(e_1 - e_3) e^{i\frac{\zeta}{2}} \sin
\frac{\zeta}{2} \\
\end{array}
\ee
The solutions are $k$-dependent and must be evaluated for the
quantized values of $k$.  Referring to the
quantization condition Eq.~(\ref{quant1}) we find that the phase $\zeta =
m \pi$ where $m$ is the longitudinal
mode index of the state.  Hence
\be
\begin{array}{ll}
E_{(++)},E_{(+-)} \propto \cos \frac{\zeta}{2} = 0 & \quad m = 1, 3,
5, \ldots \\
E_{(-+)},E_{(--)} \propto \sin \frac{\zeta}{2} = 0 & \quad m = 0, 2, 4, \ldots
\end{array}
\ee
Thus the quasi-degeneracy of the solutions is two for the bow-tie,
the solutions with identical parity under
$\sigma_Z$ form the doublets (see inset Fig.~\ref{fig:spectrum}(b)),
and these two parity types alternate in
the spectrum every free spectral
range.  Note that while we
have illustrated the analysis for for the ground state $(l=0)$ one
finds exactly the same result for the $l^{th}$
transverse mode, with doublets paired according to the index $m$,
independent of $l$.

This illustrates a general procedure, valid for any stable PO.
First, one finds by
the parabolic equation method a non-symmetrized approximate solution
$E(X,Z)$ localized on the PO.  Second, one generates the symmetrized
solutions from knowledge of the irreducible representations of the
symmetry group.  Third, one evaluates these
solutions for the quantized values of $k$; the non-zero solutions
give one the quasi-degeneracy and the
symmetry groupings (e.g. $(++)$ with $(+-)$ in the above case). The
same principles apply to
mirror resonators with the same symmetry group. Note that in the case
of a high symmetry resonator (or mirror arrangement) e.g a square or
a hexagon, for which there
exist two dimensional irreducible
representations, exact degeneracy is possible and can be found by
these methods.

\subsection{Simple Rule for Quasi-Degeneracy}

\label{section:sym2}
Although the construction just presented allows one to find the
quasi-degeneracy and symmetry pairing, it is
convenient when possible
to have a simple rule to get the quasi-degeneracy and
symmetry-pairing from the geometry of the orbit.
The quasi-degeneracy is easily determined by the following rule:

{\it The quasi-degeneracy D is equal to the number of distinct
classical periodic orbits which are related by
the spatial symmetry group and time-reversal symmetry.}

In this rule ``distinct'' orbits are defined as orbits which cannot
be mapped into one another by
time translation.  Therefore a self-retracing orbit such as the
Fabry-Perot, two-bounce orbit, only
counts as one orbit and is non-degenerate (see table ~\ref{tablesym}).
In contrast for a circulating orbit like the diamond no translation
in time will take the orbit into its time-reversed
partner.  This rule can be obtained from semiclassical methods
similar to the Gutzwiller Trace formula \cite{robbins1}.  The density of states
can be expressed by a summation over periodic orbits and their
repetitions; for stable periodic orbits
the summation over repetitions yields a delta function at the
semiclassical energies (corresponding
to the same wavevectors as we find from our quantization rule).  This
approach would give
an alternative derivation of our results which is less familiar in
optics than the parabolic equation method we have
chosen.  However the semiclassical method makes it clear that there
will be a mode for every distinct symmetry-related
PO using the definition we have just given (of course in this method,
as in the parabolic equation method, one would predict
an exact degeneracy instead of the quasi-degeneracy we have discussed).

Let us illustrate the application of this rule.  The bow-tie orbit
goes into itself under all the reflection
symmetries and so spatial symmetry generates no new orbits; however
time-reversal changes the sense
of traversal of each leg of the orbit and does give a distinct orbit.
Thus the predicted quasi-degeneracy
is two, which we found to be correct by our explicit construction
above.  In contrast, the triangle orbit
(see table~\ref{tablesym}) has a symmetry related distinct orbit and a definite
sense of circulation which is reversed
by time-reversal, hence it should have a quasi-degeneracy $2 \times 2
= 4$ leading to quartets instead of
doublets.  A few different cases of this rule are illustrated in
table~\ref{tablesym}.

The rule we have just given tells one the quasi-degeneracy, D, but
not the symmetry-pairing.  For the
case of reflection symmetries one can state a second rule which
determines these pairings.
First fold the PO of interest back into
the symmetry-reduced resonator \cite{robbins1} (see table~\ref{tablesym}) using
reflection until it
completes {\it one period in the
reduced resonator}.  The symmetry-reduced
resonator has boundaries which correspond to lines of reflection
symmetry in the original problem.  Anti-symmetric
solutions with respect to each of these lines of symmetry correspond
to Dirichlet boundary conditions; symmetric solutions must have zero
derivative corresponding
to Neumann boundary conditions.  The boundary conditions at the true boundary
of the resonator don't affect the symmetry
pairing.  For each symmetry choice one
can evaluate the phase accumulated in the reduced resonator at each
bounce, assigning a phase shift $\pi$
to each bounce off a``Dirichlet" internal boundary, and zero phase
shift for each bounce off a ``Neumann"
internal boundary.  If two symmetry types lead to the same final
phase shift (modulo $2\pi$) then those
two symmetry types will be paired and quasi-degenerate, otherwise
not.  A subtle issue is the question of how to count
bounces at the corner between two boundaries.  The answer is that the
semiclassical method really sums over orbits
nearby the PO which will then hit both boundaries and experience the
sum of the two phase shifts.

\begin{table*}[t]
\begin{center}
\begin{tabular}{|c|c|c|c|c|c|}\hline
       & symmetry- &time-   &quasi-     & symmetry-     &symmetry-   \\ 
orbit  &related    &reversal&degeneracy & reduced orbit & pairing   \\\hline
&	&	&	& &\\[-2.ex]
\includegraphics[width=1cm]{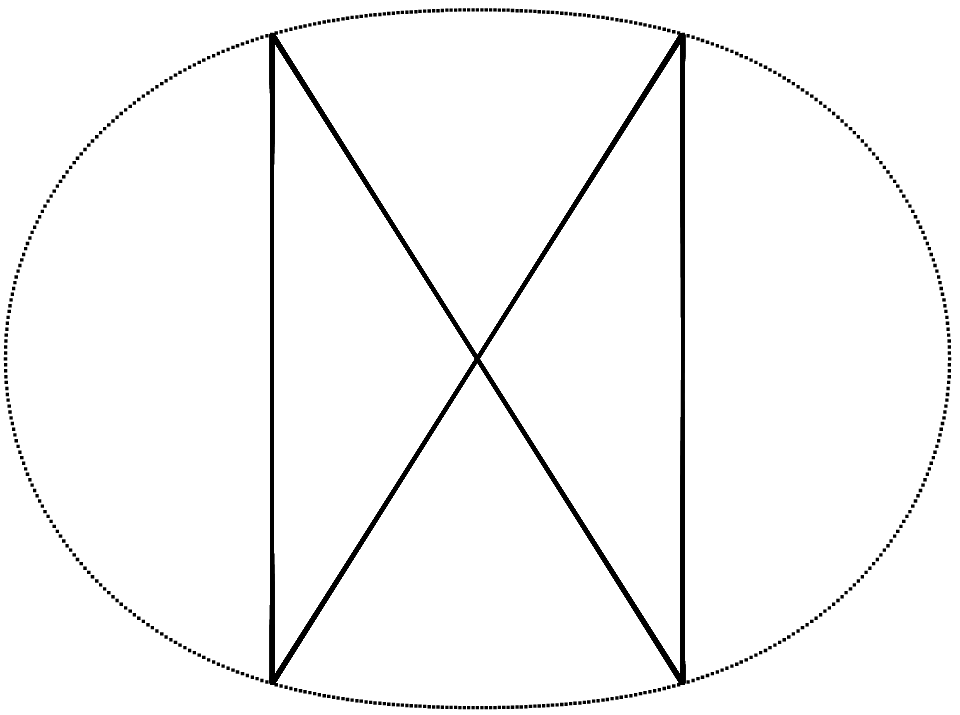}	&\raisebox{0.5ex}[-1.5ex]{1}&\raisebox{0.5ex}[-1.5ex]{2}&\raisebox{0.5ex}[-1.5ex]{2}&
\includegraphics[width=1cm]{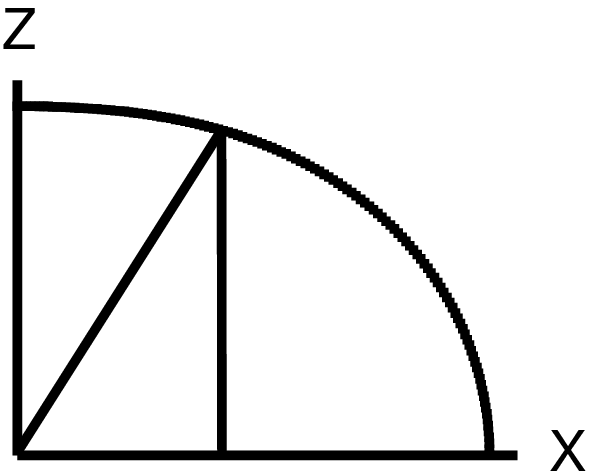}&
\raisebox{0.5ex}[-1.5ex]{$[(++), (+-)]; [(--), (-+)]$}\\
\raisebox{0.5ex}[-1.5ex]{bowtie}&	&	&	& &\\ \hline
&	&	&	& &\\[-2.ex]
\includegraphics[width=1cm]{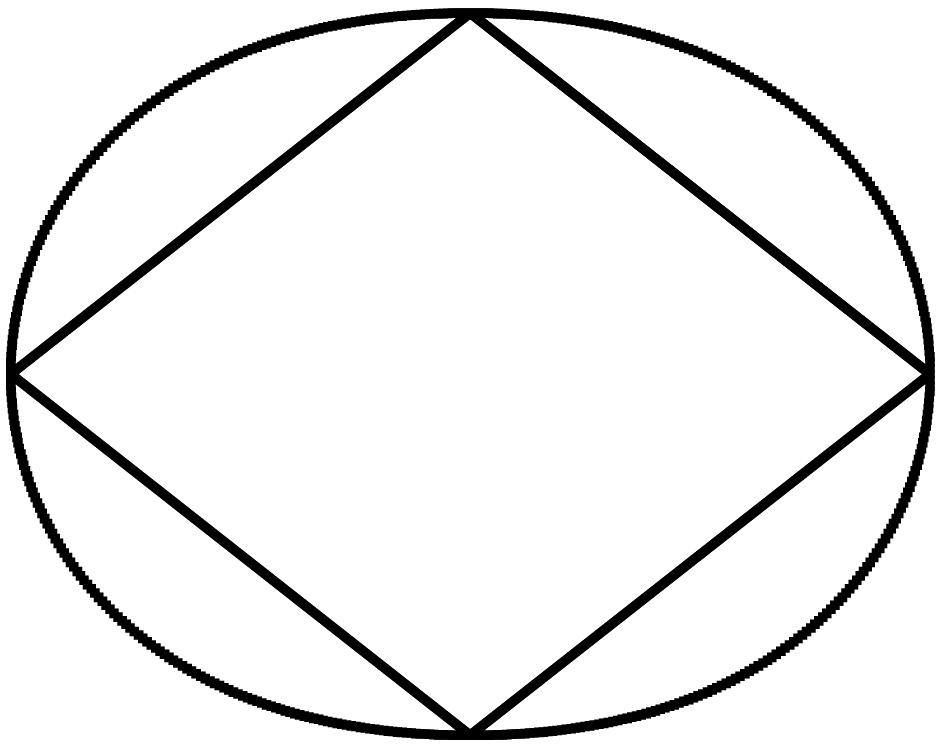}	&\raisebox{0.5ex}[-1.5ex]{1}&\raisebox{0.5ex}[-1.5ex]{2}&\raisebox{0.5ex}[-1.5ex]{2}&
\includegraphics[width=1cm]{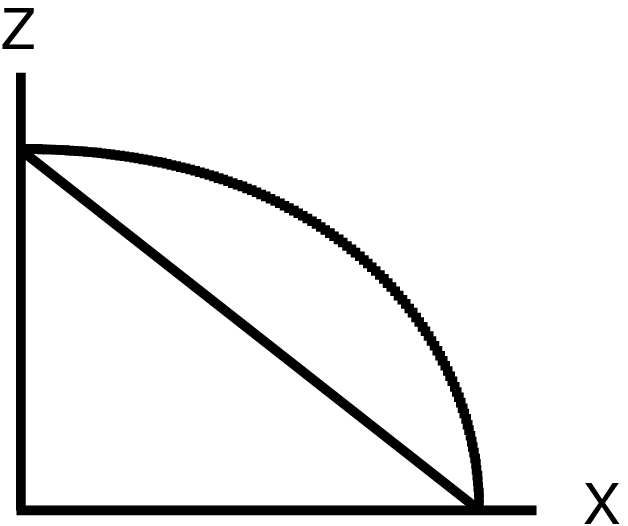}&
\raisebox{0.5ex}[-1.5ex]{$ [(++), (--)]; [(+-), (-+)]$ }\\
\raisebox{0.5ex}[-1.5ex]{diamond}&	&	&	& &\\\hline
&	&	&	& &\\[-2.ex]
\includegraphics[width=1cm]{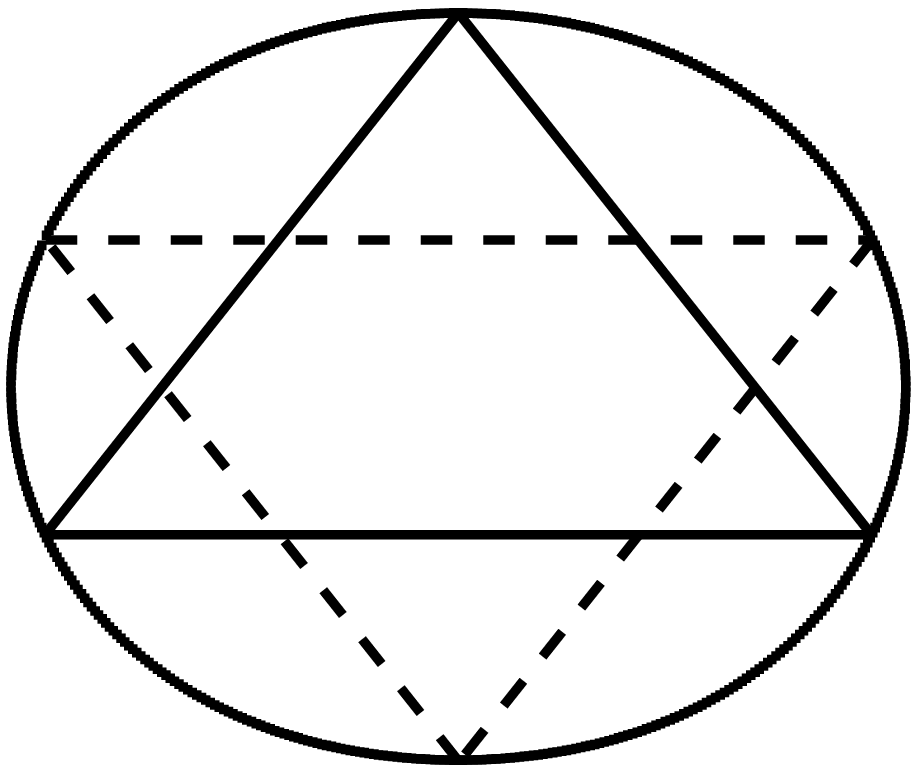}	&\raisebox{0.5ex}[-1.5ex]{2}&\raisebox{0.5ex}[-1.5ex]{2}&\raisebox{0.5ex}[-1.5ex]{4}&
\includegraphics[width=1cm]{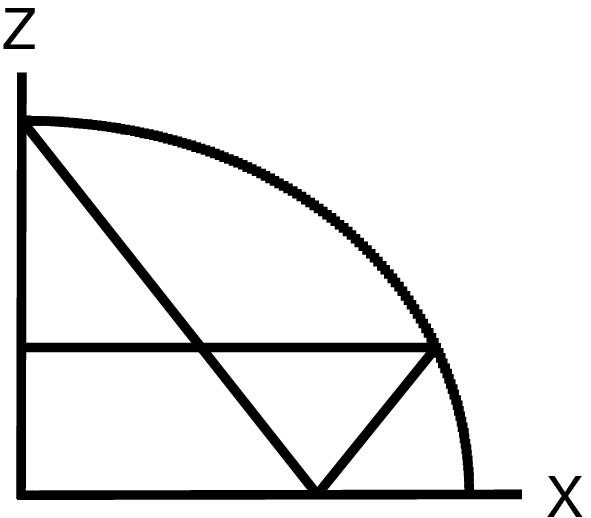}&
\raisebox{0.5ex}[-1.5ex]{$[(++), (+-), (-+), (--)]$}\\
\raisebox{0.5ex}[-1.5ex]{triangle}&	&	&	& &\\\hline
&	&	&	&  &\\[-2.ex]
\includegraphics[width=1cm]{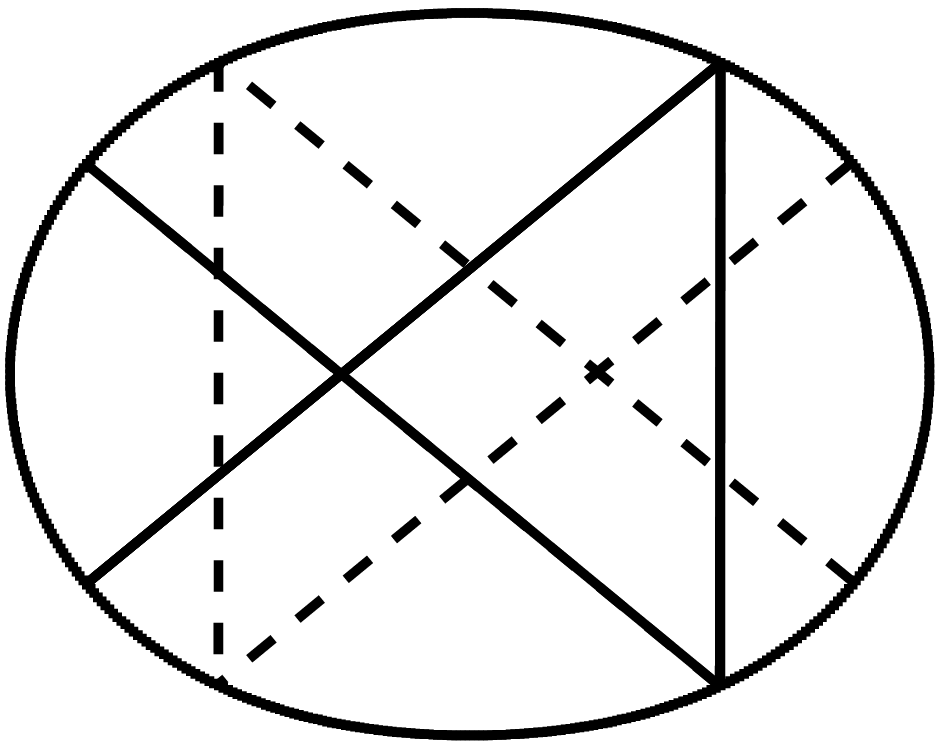}	&\raisebox{0.5ex}[-1.5ex]{2}&\raisebox{0.5ex}[-1.5ex]{1}&\raisebox{0.5ex}[-1.5ex]{2}&\includegraphics[width=1cm]{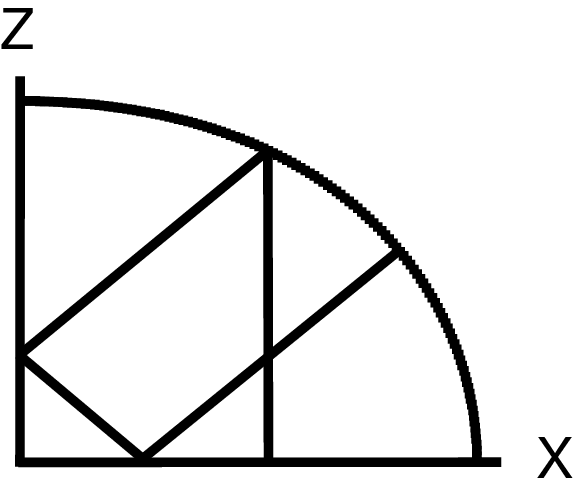}&
\raisebox{0.5ex}[-1.5ex]{$[(++), (-+)]; [(+-), (--)]$}
\\
\raisebox{0.5ex}[-1.5ex]{fish}&	&	&	& &\\\hline
&	&	&	& &\\[-2.ex]
\includegraphics[width=1cm]{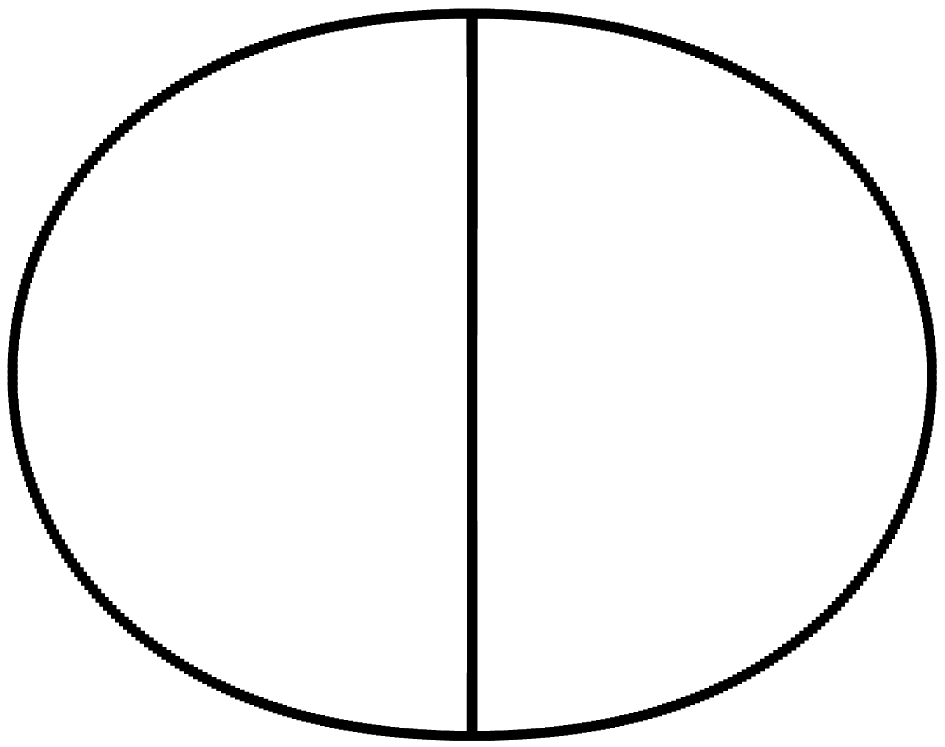}	&\raisebox{0.5ex}[-1.5ex]{1}&\raisebox{0.5ex}[-1.5ex]{1}&\raisebox{0.5ex}[-1.5ex]{1}&\includegraphics[width=1cm]{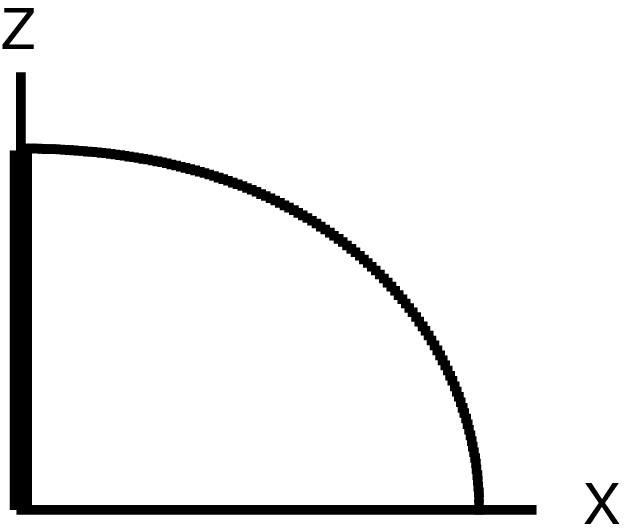}&
\raisebox{0.5ex}[-1.5ex]{$[(++)]; [(+-)]$}
\\
\raisebox{0.5ex}[-1.5ex]{fabry perot}&	&	&	& &\\\hline
\end{tabular}
\end{center}
\caption{Table illustrating the application of the two symmetry rules to
five simples POs.  The second column is the total number of orbits of this
shape related by spatial symmetry; the third column is the total 
number of orbits
of this shape generated by time-reversal symmetry.  By Rule 1 the 
quasi-degeneracy
is the product of these two numbers.  Column 4 gives the symmetry-reduced
orbit which leads via Rule 2 to the symmetry pairing indicated in column 5.}
\label{tablesym}
\end{table*}
We will illustrate this
rule for the case of the bow-tie in the quadrupole.  The symmetry
reduced PO is shown in the last column of table~\ref{tablesym}.
It has one corner bounce, one bounce on the $X$ axis and two boundary
bounces.  The boundary bounces don't matter as
they will give the same phase shift for all symmetry types.  The $X$
axis bounce will give phase shift $0$ for the
$+$ symmetry of $\sigma_Z$ and $\pi$ for the $-$ symmetry.  The
corner bounce sums the two shifts and gives:
$(+,+) \rightarrow 0,(+,-) \rightarrow \pi, (-,+) \rightarrow \pi,
(-,-) \rightarrow 2\pi$.  Adding these two
shifts modulo $2 \pi$ gives $[(+,+),(+,-)] \rightarrow 0,[(-,-),(-,+)]
\rightarrow \pi$ corresponding to the
symmetry pairing we found above.
In table~\ref{tablesym}, these two rules
are applied to a number of relevant orbits in the quadrupole.  It
should be emphasized however that the group-theoretic
projection method combined with the quantization rule which we
illustrated in this section~\ref{section:sym2} will work for any symmetry group and the
rules that we have stated are just useful shortcuts.

\subsection{Evaluation of Mode Splittings}

\label{section:split}
The symmetry analysis above can only determine the existence of
quasi-degenerate multiplets
with small splittings, it cannot estimate the size of these
splittings.  In the phase space picture
the splittings we are discussing come from tunneling between distinct
periodic orbits, referred to
as `` dynamical tunneling'' in the quantum chaos literature
\cite{DavisHeller}. Techniques have
been developed in that context for evaluating the splittings and we
now apply those to the system
we are considering.

Dynamical tunneling in integrable systems is essentially similar to
the textbook example of tunneling in one dimension
(note that conservative dynamics in 1D is always
integrable), and the splittings in such a case come from first order
perturbation theory in the tunneling matrix
element through an effective barrier, just as they do for the
one-dimensional double well potential.
Systems of the type we are considering, with mixed dynamics however
show a very striking difference. It has been
found by both numerical simulations and analytic arguments that the
dynamical tunneling splittings in mixed
systems are typically many order of magnitude larger than found for
similar but integrable systems \cite{ullmo1,frischat} (e.g in
quadrupole vs. elliptical billiards). This difference  can be traced
to the mechanism of
``chaos-assisted tunneling'' (CAT) . As opposed
to the ``direct" processes when the particle
(ray) ``tunnels" directly from one orbit to the other, the CAT
corresponds to the following  three-step process: (i) tunneling
from the periodic orbit to the nearest point of the chaotic ``sea",
(ii) {\it classical} propagation in the chaotic
portion of the phase space until the neighborhood of the other
periodic orbit is reached, (iii) tunneling from the chaotic
sea to the other periodic orbit. Note that  the chaos-assisted
processes are formally of higher order in the
perturbation theory. However the corresponding matrix elements are
much larger than those of the direct process. This
can be understood intuitively as the tunneling from the periodic
orbit to the chaotic sea typically involves a
much smaller  ``violation" of classical mechanics and therefore has
an exponentially
larger amplitude.

\begin{figure}[h]
\centering
\includegraphics[width=0.6\linewidth]{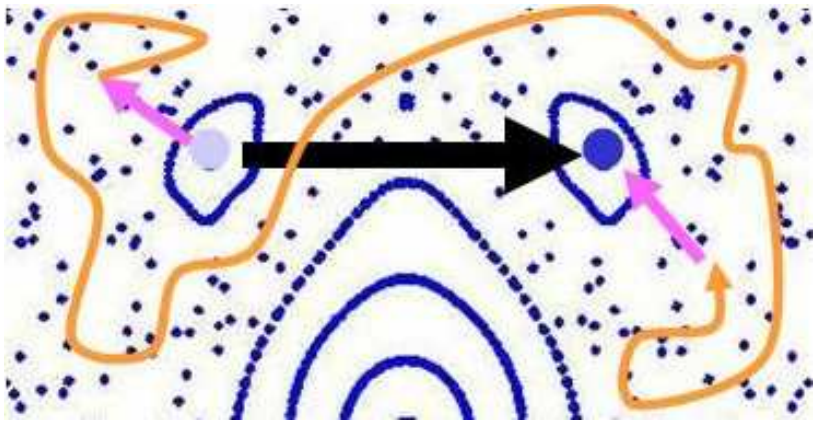}
\caption{Schematic indicating a direct tunneling process (black arrow)
and a chaos-assisted tunneling process (yellow arrow) which would contribute
to splitting of bow-tie doublets.}
\label{fig:schemCAT}
\end{figure}

The contribution of chaos-assisted tunneling can be evaluated both
qualitatively and quantitatively using the so called
``three-level model" \cite{ullmo1,ullmo2}, where the chaotic energy levels
(the eigenstates localized in the chaotic portion of the phase space)
are represented by a single state $E_C$ with known
statistical properties. The straightforward diagonalization of the
resulting matrix yields \cite{ullmo2}
\begin{eqnarray}
\Delta E_{\rm CAT} & \simeq & \frac{\left| V_{RC} \right|^2 }{E_R - E_C}
\label{eq:dE_cat}
\end{eqnarray}
where $E_R$ is the semiclassical energy of the ``regular" states
(localized at the periodic orbits)
     which does not include the tunneling contribution, and $V_{RC}$ is the
corresponding coupling matrix element with the chaotic state.
The resonant denominator in Eq. (\ref{eq:dE_cat}) leads to strong
fluctuations of the
CAT-related doublets, as is found in numerical simulations\cite{ullmo2}.
The average behavior of the splittings however is determined by the
matrix element $V_{RC}$.
By virtue of the Wigner transformation \cite{gutz_book} it can be shown
\cite{evgeni1} that
$\left| V_{RC} \right|^2$ is proportional to the overlap of the
Wigner transforms of the "regular" and ``chaotic" states:
\begin{eqnarray}
\left| V_{RC} \right|^2 \propto \int d\phi \int d\sin\chi \
W_C\left(\phi, \sin\chi\right)
W_R\left(\phi, \sin\chi\right)
\label{eq:VRC2}
\end{eqnarray}
Assuming that, as required by Berry's conjecture\cite{berry1}, {\it on
the average} the Wigner function of a
chaotic state is equally distributed across the chaotic portion of
the phase space, and using the analytical
expressions for the regular eigenstates calculated earlier, we find
\begin{eqnarray}
\langle \Delta E_{\rm CAT} \rangle & \propto & \exp\left( - {\cal A}
k R_0 \right)
\label{eq:dE_area}
\end{eqnarray}
where ${\cal A}$ is the area in the Poincare Surface of Section (in
$(\phi, \sin\chi)$ coordinates) occupied by the
stable island supporting the regular eigenstate. Note that Eq.
(\ref{eq:dE_area}) holds only on average, since
chaos-assisted tunneling always leads to strong fluctuations of the
splittings which are of the same order as
the average \cite{ullmo1}.

\begin{figure}[h]
\centering
\includegraphics[width=0.7\linewidth]{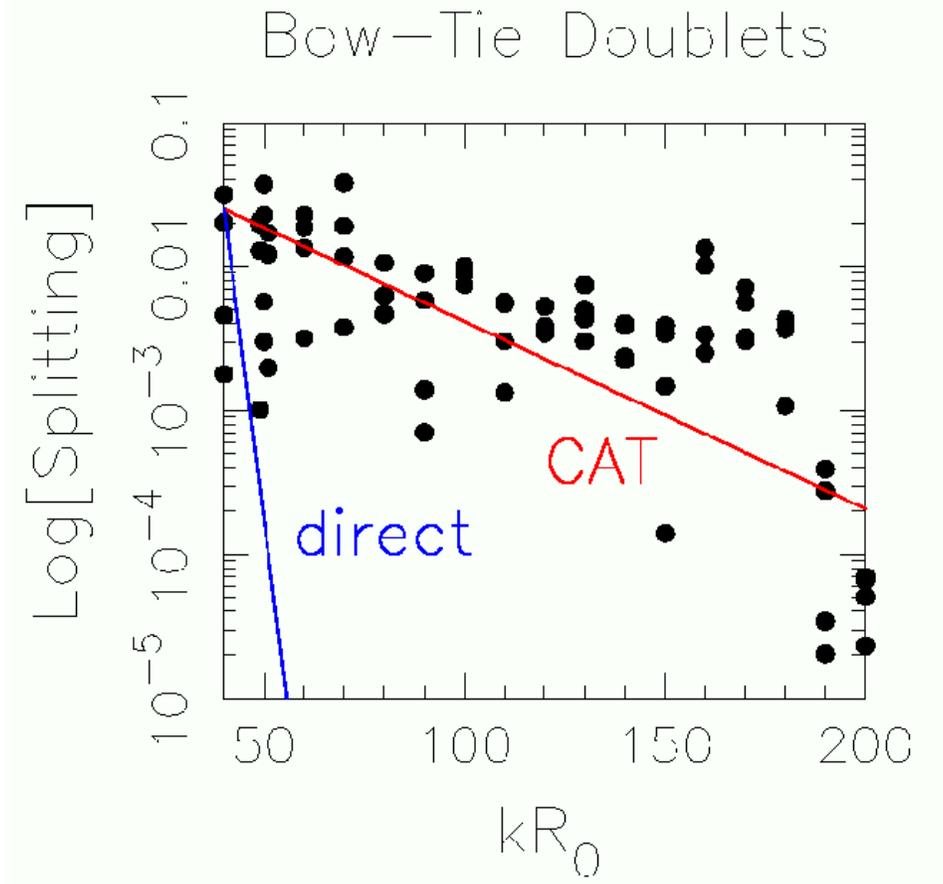}
\caption{The numerically determined splittings of bow-tie doublets
for a closed quadrupole resonator with $\epsilon = 0.14$ (black dots)
vs kR; the red line denotes the prediction of Eq.~(\ref{eq:dE_area}) for the average
splitting, the blue line an estimate of the splitting based on the
``direct'' coupling.  Note the large enhancement due to chaos-assisted
tunneling and the large fluctuations around the mean splitting.}
\label{fig:CAT}
\end{figure}

In Fig. \ref{fig:CAT} we show the numerically calculated splittings
for the bow-tie resonances in
a resonant cavity with a fixed quadrupolar deformation $\epsilon =
0.14$, for different values of $kR$.
Note that although there are large fluctuations in the numerical data
(as previously noted), the data are consistent with Eq.
(\ref{eq:dE_area}), while a
calculation based on the ``direct" coupling severely
underestimates the splittings.  Unfortunately, due to the large
fluctuations in the splittings, knowledge of the average splitting
size does not accurately predict the splitting of a specific doublet.
Note also that small violations of symmetry in the fabrication of
the resonator may lead to much larger splittings then these
tunnel splittings; such an effect was recently observed for triangle-based
modes of GaAs ARC micro-lasers \cite{gmachl1}.

For the bow-tie orbit the quasi-degeneracy is associated with
time-reversal symmetry, but as discussed above, it can
equally well be associated with spatial symmetries and indeed this
was the situation in the first work on dynamical
tunneling \cite{DavisHeller}. Splittings of this type can also be
adequately described using the framework developed in the present section.

\section{Conclusions}

We have generalized gaussian optical resonator theory to describe
the resonances associated with stable periodic orbits of arbitrary
shaped two-dimensional dielectric resonators using the parabolic
equation method.  The
correspondence to ray optics emerges naturally when imposing the
boundary conditions
at the dielectric interface which leads to the appropriate ABCD matrices
for reflection, transmission and propagation.  For a 
perfectly-reflecting cavity one
gets quantized solutions at real values of $k$ localized around the PO
with mode spacings given by $\Delta k_{long} = 2 \pi /L,
\Delta k_{trans} = \varphi /L$ where $L$ is the length of the PO and
$\varphi$ is the Floquet phase associated with the eigenvalues of the
monodromy matrix (round-trip ABCD matrix).  For a dielectric cavity one finds
similarly localized quasi-bound solutions at quantized complex values of $nk$;
in this case the imaginary part of $nk$ is determined by the Fresnel refractive
loss at each bounce of the PO.  Within this approximation the mode spacings
are unchanged from the closed case (except for the trivial factor of $n$).
These regular modes coexist in a generic resonator with more
complicated modes associated with the chaotic regions of phase space.
Generalization of our results to the three-dimensional case appears
straightforward for the scalar case and one expects only to have
three-dimensional versions of the ABCD matrices enter the theory leading
to some difference in details.  More interesting would be the inclusion
of the polarization degree of freedom, which seems possible in principle,
but which we haven't explored as yet.

We noted that for a cavity with discrete symmetries one will typically
be able to construct several symmetry-related, nominally degenerate
solutions of the
wave equation for each PO.  However group-theoretic arguments indicate
that the these solutions cannot be exactly degenerate and lead us to construct
symmetrized solutions which form quasi-degenerate multiplets.  We presented
a construction and then two simples rules for calculating the quasi-degeneracy
of these multiplets and their symmetry quantum numbers.  In the final section
of the paper we show that the splittings of these multiplets are much larger
than expected, due to the phenomenon of ``chaos-assisted'' tunneling, and
estimate the average splitting in terms of classical quantities.

There are several limitations of this work which we hope to address in
future work.  One obvious shortcoming is the prediction of zero width
modes for the dielectric cavity if the underlying PO has all of its
bounces above total internal reflection.  Internal reflection is not
perfect for these systems at any angle of incidence for two reasons.
First, as these solutions describe gaussian beams with some momentum spread,
every solution should have some plane-wave amplitude at an angle of
incidence for which it can be partially transmitted.  This type of
correction exists even for a gaussian beam incident on an infinite
planar surface and leads to an outgoing beam direction which can be
significantly different from that predicted from Snell's law; we have analyzed
the beam deflection for this case recently
\cite{tureci1}.   For the case of the bow-tie modes of the
dielectric resonator the effect
of the momentum spread in inducing a finite width was also evaluated using a
semiclassical method in ref. \cite{evgeni2}.  Second, due to the 
curvature of the
interface in such resonators, there will always be some evanescent
leakage, which we
can think of as due to direct tunneling through the angular momentum
barrier as is
known to occur even for perfectly circular resonators.   We have evaluated
this type of correction also recently \cite{evgeni1}. It is still not
clear to what extent these effects can be accounted for systematically within a
generalization of our approach here, e.g. to higher order in $kl$.
One possibility we are exploring is that a generalized ray optics with
non-specular effects included can describe the open resonator and its
emission pattern.  Both experiments \cite{science,rex1} and
numerical studies \cite{rex1,tureci1} demonstrate that for $kl$ not too
large ($ \sim
50-100$) these higher order effects must be taken into account.

This work was partially supported by NSF grant DMR-0084501.


\begin{thebibliography}{}
\newcommand{\enquote}[1]{``#1''}
\expandafter\ifx\csname url\endcsname\relax
  \def\url#1{{#1}}\fi
\expandafter\ifx\csname urlprefix\endcsname\relax\def\urlprefix{}\fi
\bibitem{campillo_book}
R.~K. Chang, A.~K. Campillo, eds., {\em Optical Processes in
  Microcavities\/} (World Scientific, Singapore, 1996).

\bibitem{yokoyama1}
H.~Yokoyama, \enquote{Physics and device applications of optical
  microcavities,} Science {\bf 256}, 66--70 (1992).

\bibitem{nobel}
A.~D. Stone, \enquote{Wave-chaotic optical resonators and lasers,} Phys. Scr.
  {\bf T90}, 248--262 (2001).

\bibitem{science}
C.~Gmachl, F.~Capasso, E.~E. Narimanov, J.~U. N\"ockel, A.~D. Stone, J.~Faist,
  D.~L. Sivco, and A.~Y. Cho, \enquote{High-power directional emission from
  microlasers with chaotic resonators,} Science {\bf 280}, 1556--1564 (1998).

\bibitem{nature}
J.~U. N\"ockel and A.~D. Stone, \enquote{Ray and wave chaos in asymmetric
  resonant optical cavities,} Nature {\bf 385}, 45--47 (1997).

\bibitem{campillo1}
A.~J. Campillo, J.~D. Eversole, and H.~B. Lin, \enquote{Cavity quantum
  electrodynamic enhancement of stimulated-emission in microdroplets,}
  Phys.~Rev.~Lett. {\bf 67}, 437--440 (1991).

\bibitem{lin1}
H.~B. Lin, J.~D. Eversole, and A.~J. Campillo, \enquote{Spectral properties of
  lasing microdroplets,} J.~Opt.~Soc.~Am.~B {\bf 9}, 43--50 (1992).

\bibitem{vahala1}
S.~M. Spillane, T.~J. Kippenberg, and K.~J. Vahala, \enquote{Ultralow-threshold
  Raman laser using a spherical dielectric microcavity,} Nature {\bf 415},
  621--623 (2002).

\bibitem{little1}
B.~E. Little, J.~S. Foresi, G.~Steinmeyer, E.~R. Thoen, S.~T. Chu, H.~A. Haus,
  E.~P. Ippen, L.~C. Kimerling, and W.~Greene, \enquote{Ultra-compact Si-SiO2
  microring resonator optical channel dropping filters,} IEEE Photonics
  Technol. Lett. {\bf 10}, 549--551 (1998).

\bibitem{chang3}
S.~X. Qian, J.~B. Snow, H.~M. Tzeng, and R.~K. Chang, \enquote{Lasing droplets
  - highlighting the liquid-air interface by laser-emission,} Science {\bf
  231}, 486--488 (1986).

\bibitem{poon1}
A.~W. Poon, F.~Courvoisier, and R.~K. Chang, \enquote{Multimode resonances in
  square-shaped optical microcavities,} Opt. Lett. {\bf 26}, 632--634 (2001).

\bibitem{braun1}
I.~Braun, G.~Ihlein, F.~Laeri, J.~U. N\"ockel, G.~Schulz-Ekloff, F.~Schuth,
  U.~Vietze, O.~Weiss, and D.~Wohrle, \enquote{Hexagonal microlasers based on
  organic dyes in nanoporous crystals,} Appl. Phys. B-Lasers Opt. {\bf 70},
  335--343 (2000).

\bibitem{mekis}
A.~Mekis, J.~U. N\"ockel, G.~Chen, A.~D. Stone, and R.~K. Chang, \enquote{Ray
  chaos and q spoiling in lasing droplets,} Phys. Rev. Lett. {\bf 75},
  2682--2685 (1995).

\bibitem{nockel1}
J.~U. Nockel, A.~D. Stone, G.~Chen, H.~L. Grossman, and R.~K. Chang,
  \enquote{Directional emission from asymmetric resonant cavities,} Opt. Lett.
  {\bf 21}, 1609--1611 (1996).

\bibitem{gornik1}
S.~Gianordoli, L.~Hvozdara, G.~Strasser, W.~Schrenk, J.~Faist, and E.~Gornik,
  \enquote{Long-wavelength {$\lambda=10 \mu m$} quadrupolar-shaped GaAs-AlGaAs
  microlasers,} IEEE J. Quantum Electron. {\bf 36}, 458--464 (2000).

\bibitem{chang1}
S.~Chang, R.~K. Chang, A.~D. Stone, and J.~U. N\"ockel, \enquote{Observation of
  emission from chaotic lasing modes in deformed microspheres: displacement by
  the stable-orbit modes,} J. Opt. Soc. Am. B-Opt. Phys. {\bf 17}, 1828--1834
  (2000).

\bibitem{rex1}
N.~B. Rex, H.~E. Tureci, H.~G.~L. Schwefel, R.~K. Chang, and A.~D. Stone,
  \enquote{Fresnel filtering in lasing emission from scarred modes of
  wave-chaotic optical resonators,} Phys. Rev. Lett. {\bf 88}, art.
  no.094\,102 (2002).

\bibitem{rex_thesis}
N.~B. Rex, {\em Regular and chaotic orbit Gallium Nitride microcavity
  lasers\/}, Ph.D. thesis, Yale University (2001).

\bibitem{gmachl1}
C.~Gmachl, E.~E. Narimanov, F.~Capasso, J.~N. Baillargeon, and A.~Y. Cho,
  \enquote{Kolmogorov-Arnold-Moser transition and laser action on scar modes in
  semiconductor diode lasers with deformed resonators,} Opt. Lett. {\bf 27},
  824--826 (2002).

\bibitem{sblee1}
S.~B. Lee, J.~H. Lee, J.~S. Chang, H.~J. Moon, S.~W. Kim, and K.~An,
  \enquote{Observation of scarred modes in asymmetrically deformed
  microcylinder lasers,} Phys. Rev. Lett. {\bf 88}, art. no.033903
  (2002).

\bibitem{heller1}
E.~J. Heller, \enquote{Bound-state eigenfunctions of classically chaotic
  hamiltonian-systems - Scars of periodic orbits,} Phys.~Rev.~Lett. {\bf 53},
  1515--1518 (1984).

\bibitem{harald1}
H.~G.~L. Schwefel, N.~B. Rex, H.~E. Tureci, R.~K. Chang, and A.~D. Stone,
  \enquote{Dramatic shape sensitivity of emission patterns for similarly
  deformed cylindrical polymer lasers,} CLEO/QELS 2002.

\bibitem{berry2}
M.~V. Berry, \enquote{Regularity and chaos in classical mechanics, illustrated
  by three deformations of a circular billiard,} Eur.~J.~Phys. {\bf 2}, 91--102
  (1981).

\bibitem{robnik1}
B.~Li and M.~Robnik, \enquote{Geometry of high-lying eigenfunctions in a plane
  billiard system having mixed type classical dynamics,} J.~Phys.~A {\bf 28},
  2799--2818 (1995).

\bibitem{keller1}
J.~B. Keller and S.~I. Rubinow, \enquote{Asymptotic Solution of Eigenvalue
  Problems,} Ann.~Phys. {\bf 9}, 24--75 (1960).

\bibitem{gutz_book}
M.~C. Gutzwiller, {\em Chaos in classical and quantum mechanics\/} (Springer,
  New York, USA, 1990).

\bibitem{husimi1}
S.~D. Frischat and E.~Doron, \enquote{Quantum phase-space structures in
  classically mixed systems: A scattering approach,} J. Phys. A-Math. Gen. {\bf
  30}, 3613--3634 (1997).

\bibitem{babic_book}
V.~M. Babi\v{c} and V.~S. Buldyrev, {\em Asymptotic Methods in Shortwave
  Diffraction Problems\/} (Springer, New York, USA, 1991).

\bibitem{siegman_book}
A.~E. Siegman, {\em Lasers\/} (University Science Books, Mill Valley,
  California, 1986).

\bibitem{maslov_book}
V.~P. Maslov and M.~V. Fedoriuk, {\em Semiclassical Approximations in Quantum
  Mechanics\/} (Reidel, Boston, USA, 1981).

\bibitem{chernikov1}
N.~A. Chernikov, \enquote{System whose hamiltonian is a time-dependent
  quadratic form in x and p,} Sov Phys-Jetp Engl Trans {\bf 26}, 603--608
  (1968).

\bibitem{young1}
E.~S.~C. Ching, P.~T. Leung, A.~Maassen van~den Brink, W.~M. Suen, T.~S. S.,
  and K.~Young, \enquote{Quasinormal-mode expansion for waves in open systems,}
  Rev. Mod. Phys. {\bf 70}, 1545--1554 (1998).

\bibitem{ra1}
J.~W. Ra, H.~L. Bertoni, and L.~B. Felsen, \enquote{Reflection and transmission
  of beams at a dielectric interface,} SIAM J. Appl. Math {\bf 24}, 396--413
  (1973).

\bibitem{robbins1}
J.~M. Robbins, \enquote{Discrete symmetries in periodic-orbit theory,}
  Phys.~Rev.~A. {\bf 40}, 2128--2136 (1989).

\bibitem{DavisHeller}
M.~J. Davis and E.~J. Heller, \enquote{Multidimensional wave functions from
  classical trajectories,} J.~Chem.~Phys. {\bf 75}, 246 (1981).

\bibitem{ullmo1}
O.~Bohigas, S.~Tomsovic, and D.~Ullmo, \enquote{Manifestations of classical
  phase space structures in quantum mechanics,} Phys.~Rep. {\bf 223}, 45
  (1993).

\bibitem{frischat}
S.~D. Frischat and E.~Doron, \enquote{Semiclassical description of tunneling in
  mixed systems: case of the annular billiard,} Phys.~Rev.~Lett. {\bf 75}, 3661
  (1995).

\bibitem{ullmo2}
F.~Leyvraz and D.~Ullmo, \enquote{The level splitting distribution in
  chaos-assisted tunneling,} J.~Phys.~A {\bf 29}, 2529 (1996).

\bibitem{evgeni1}
E.~E. Narimanov, unpublished.

\bibitem{berry1}
M.~V. Berry, \enquote{Regular and irregular semiclassical wavefunctions,}
  J.~Phys.~A {\bf 10}, 2083 (1977).

\bibitem{tureci1}
H.~E. Tureci and A.~D. Stone, \enquote{Deviation from Snell's law for beams
  transmitted near the critical angle: application to microcavity lasers,} Opt.
  Lett. {\bf 27}, 7--9 (2002).

\bibitem{evgeni2}
E.~E. Narimanov, G.~Hackenbroich, P.~Jacquod, and A.~D. Stone,
  \enquote{Semiclassical theory of the emission properties of wave-chaotic
  resonant cavities,} Phys. Rev. Lett. {\bf 83}, 4991--4994 (1999).
\end{thebibliography}
\end{document}